\documentclass[prb,aps,twocolumn,amsmath,amssymb]{revtex4}

\usepackage{epsfig}% Include figure files
\usepackage{graphicx}% Include figure files
\usepackage{dcolumn}% Align table columns on decimal point
\usepackage{bm}% bold math
\usepackage{color}
\usepackage{subfigure}

\begin{document}

\title{Angle resolved photoemission spectroscopy reveals spin charge separation in metallic MoSe$_2$ grain boundary}
\author{Yujing Ma,$^1$ Horacio Coy Diaz,$^1$ Jos\'e Avila,$^ {2,3}$ Chaoyu Chen,$^ {2,3}$ Vijaysankar Kalappattil,$^1$ Raja Das,$^1$\\ 
Manh-Huong Phan,$^1$ Tilen \v{C}ade\v{z}, $^{4,5}$ Jos\'e M. P. Carmelo,$^{5,6,4*}$
Maria C. Asensio,$^{2,3,*}$ and Matthias Batzill,$^{1*}$}

\affiliation{$^{1}$ Department of Physics, University of South Florida, Tampa, FL 33620, USA}
\affiliation{$^{2}$ Synchrotron SOLEIL, L'Orme des Merisiers, Saint Aubin-BP 48, 91192 Gif sur Yvette Cedex, France}
\affiliation{$^{3}$ Universit\'e Paris-Saclay, L'Orme des Merisiers, Saint Aubin-BP 48, 91192 Gif sur Yvette Cedex, France}
\affiliation{$^{4}$ Beijing Computational Science Research Center, Beijing 100193, China}
\affiliation{$^{5}$ Center of Physics of University of Minho and University of Porto, P-4169-007 Oporto, Portugal}
\affiliation{$^{6}$ Department of Physics, University of Minho, Campus Gualtar, P-4710-057 Braga, Portugal}

\begin{abstract}
*e-mail: mbatzill@usf.edu; asensio@synchrotron-soleil.fr; carmelo@fisica.uminho.pt\\ \\
{\bf  Material line defects are one-dimensional structures but the search and proof of electron behaviour consistent with the reduced dimension of such defects has been so far unsuccessful. Here we show using angle resolved photoemission spectroscopy that twin-grain boundaries in the layered semiconductor MoSe$_2$ exhibit parabolic metallic bands. The one-dimensional nature is evident from a charge density wave transition, whose periodicity is given by $k_{\rm F}/\pi$, consistent with scanning tunneling microscopy and angle resolved photoemission measurements. Most importantly, we provide evidence for spin- and charge-separation, the hallmark of one-dimensional quantum liquids. Our studies show that the spectral line splits into distinctive spinon and holon excitations whose dispersions exactly follow the energy-momentum dependence calculated by a Hubbard model with suitable finite-range interactions. Our results also imply that quantum wires and junctions can be isolated in line defects of other transition metal dichalcogenides, which may enable quantum transport measurements and devices.} 
\end{abstract}

\maketitle

1D electron systems (1DES) are sought for their potential applications in novel quantum devices as well as for enabling fundamental scientific discoveries in materials with reduced dimensions. Certainly, 1D electron dynamics plays a central role in nanoscale materials physics, from nanostructured semiconductors to (fractional) quantum Hall edge states\cite{Wen-90,Stone-91}. Furthermore, it is an essential component in Majorana fermions\cite{Mourik-12,Kitaev-01} and is discussed in relation to the high-$T_{\rm c}$ superconductivity mechanism\cite{Orgad-01}. However, truly 1D quantum systems that permit testing of theoretical models by probing the full 
momentum-energy $(k,\omega)$-space are sparse and consequently angle-resolved photoelectron spectroscopy (ARPES) measurements have only been possible on quasi-1D materials consisting of 2D- or 3D-crystals that exhibit strong 1D anisotropy\cite{Ohtsubo-15,Segovia-05,Claessen-02,Sing-03,Zwick-98}. 

Electrons confined in one-dimension (1D) behave fundamentally different from the Fermi-liquid in higher dimensions\cite{Haldane-81,Voit-95,Glazman-12}. While there exist various quasi-1D materials that have strong 1D anisotropies and thus exhibit 1D properties, strictly 1D metals, i.e. materials with only periodicity in 1D that may be isolated as a single wire, have not yet been described as 1D quantum liquids. Grain boundaries in 2D van der Waals materials are essentially 1D and recent DFT simulations on twin grain boundaries in MoS$_2$\cite{Zou-13} and MoSe$_2$\cite{Lehtinen-15} have indicated that those defects should exhibit a single band intersecting the Fermi level. Therefore, such individual line defects are exceptional candidates for truly 1D metals. 

In the case of quasi-1D Mott-Hubbard insulators (MHI)\cite{Schlappa-12,Kim-06,Auslaender-05,Kim-96}, 
there is strong evidence for the occurrence of the so called spin-charge separation\cite{Kim-06,Auslaender-05}. 
Recently, strong evidence of another type of separation in these quasi-1D compounds was found, specifically
a spin-orbiton separation with the orbiton carrying an orbital excitation\cite{Schlappa-12}. 

The theoretical treatment of MHI is easier compared to that of the physics of 1D metals. The ground state of a MHI has no holons and no spinons 
and the dominant one-electron excited states are populated by one holon and one spinon, as defined by the Tomonaga 
Luttinger liquid (TLL) formalism\cite{Voit-95}. 
For 1DES metals the scenario is however more complex, as the holons are present in both the ground and the excited states. 
Zero spin-density ground states have no spinons. Consequently, the experimental verification of key features of 1DES, especially 
the spin-charge separation, remains still uncertain\cite{Ohtsubo-15,Losio-01,Ahn-03,Weitering-11}. 

The theoretical description of 1DES low-energy excitations in terms of spinons and holons, based on the TLL formalism, has been a corner stone of 
1D electron low-energy dynamics\cite{Voit-95}. The rather effective approximation of the relation of energy versus momentum in 1D fermions by a strictly linear dispersion relation, makes the problem accessible and solvable, by calculating analytically the valuable many-body low-energy dynamics of the system. This drastic assumption has provided an effective tool to describe low-energy properties of 1D quantum liquids in terms of quantized linear collective sound modes, named spinons (zero-charge spin excitations) and holons (spinless charge excitations), respectively. 
However, this dramatic simplification is only valid in the range of low-energy excitations, very close to the Fermi level.

More recently, sophisticated theoretical tools have been developed that are capable to extend this description to high-energy excitations away from the Fermi-level\cite{Glazman-12,Benthien-04,Carmelo-05,Carmelo-06,Carmelo-08,Carmelo-17,Glazman-09}. Particularly, the pseudofermion dynamical theory (PDT)\cite{Carmelo-05,Carmelo-06,Carmelo-08,Carmelo-17} allows to compute one-particle spectral functions in terms of spinon and holon features, in the full energy versus momentum space ($(k,\omega)$-plane). The exponents controlling the low- and high-energy spectral-weight distribution are functions of momenta, differing significantly from the predictions of the TLL if applied to the high-energy regime\cite{Benthien-04,Carmelo-05,Carmelo-06,Carmelo-08,Carmelo-17}. To the best of our knowledge, while other theoretical approaches, beyond the TLL limit, have also been recently developed\cite{Glazman-12,Glazman-09}, no direct photoemission measurements of spin-charge separation in a pure metallic 1DES has been reported so far. Even more important, a  theoretical 1D approach with electron finite-range interactions entirely consistent with the photoemission data in the full energy versus momentum space has never been reported before \cite{Haldane-81,Schulz-90,Voit-95}.

Here we present a description of the non-Fermi liquid behaviour of a metallic 1DES with suitable finite-range interactions over the 
entire $(k,\omega)$-plane that matches the experimentally determined weights
over spin- and charge- excitation branches. This has been accomplished by carrying out the first ARPES study of a 1DES hosted in an intrinsic line defect of a material and by developing a new theory taking electron finite-range interactions within an extended 1D Hubbard model into account.

\begin{figure}
\centerline{\includegraphics[width=10cm]{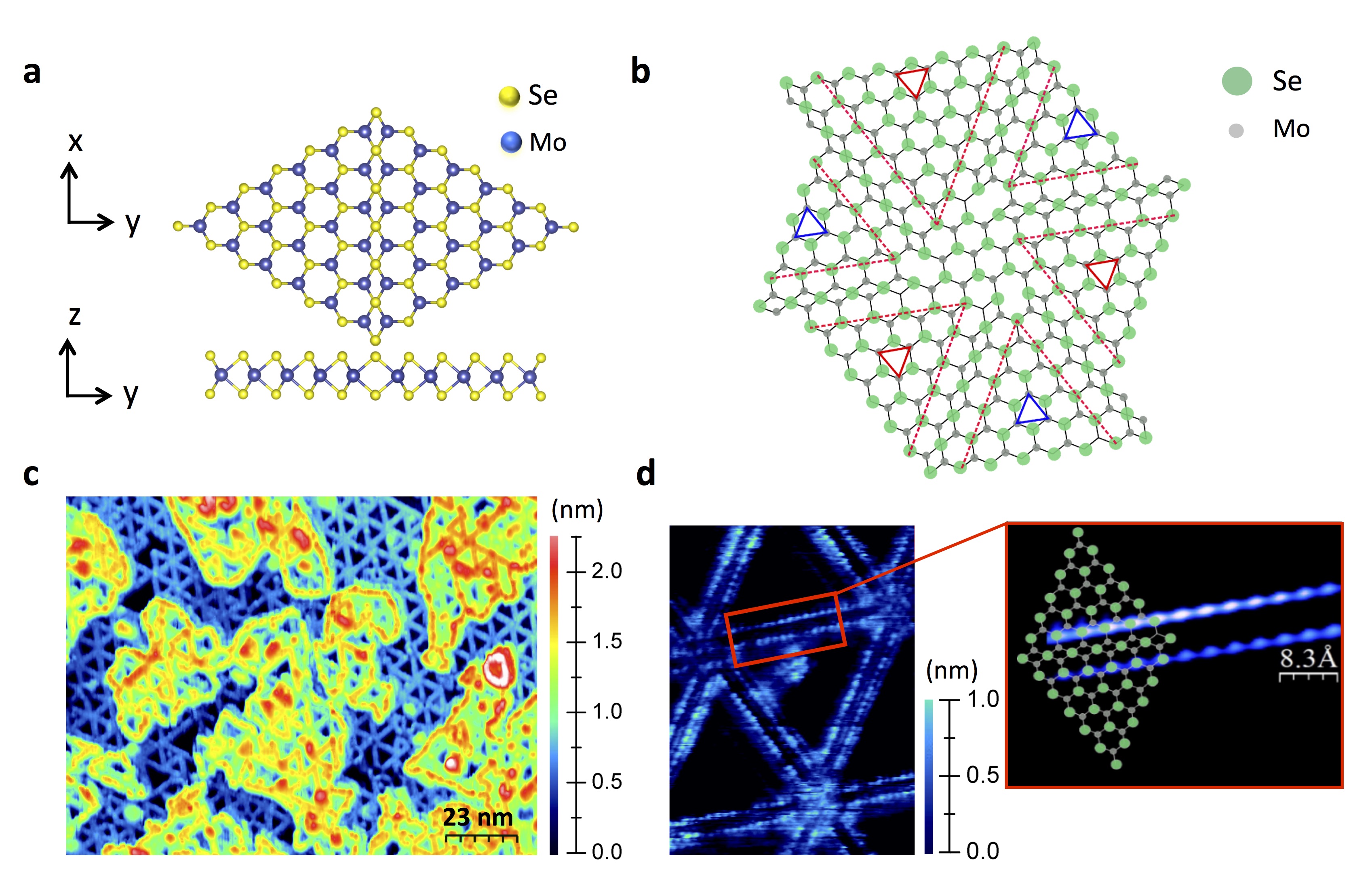}}
\caption{{\bf Defect structure of mirror twin grain boundaries (MTBs) in monolayer MoSe$_2$.} 
(a) Ball-and-stick model of a MTB, 
indicating that the grain boundary is Se deficient. (b) Arrangements of the three equivalent MTB directions gives rise to 
a cross-hatched grain boundary network. (c) Large-scale (150 nm $\times$ 110 nm) STM image of 1-2 monolayers of 
MoSe$_2$ grown by MBE on MoS$_2$. The MTBs appear as bright lines forming a dense network of aligned line defects. In 
higher resolution images shown in (d) the defect lines appear as two parallel lines. Imaging at room temperature allows 
resolving atomic corrugation along these lines that are attributed to atom positions in the Se-rows adjacent to the defect 
line, as the overlay of the model illustrates.}
\label{figure1}
\end{figure}

\begin{figure}
\centerline{\includegraphics[width=9cm]{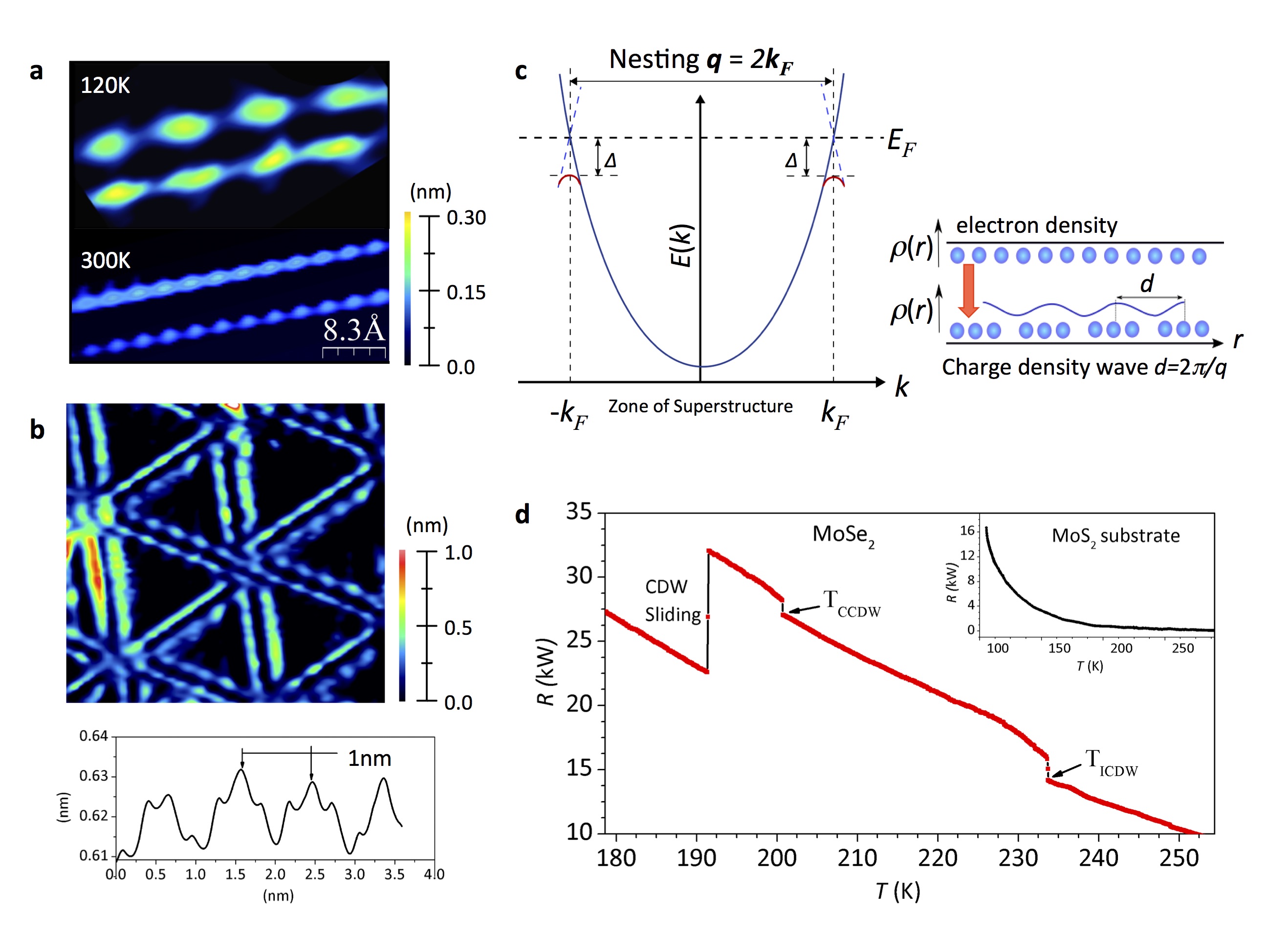}}
\caption{{\bf Charge density wave (CDW) transition in MTBs.} 
(a) STM images of a single MTB at low temperatures (120 K) exhibit three times the periodicity than the atomic corrugation imaged at room temperature. In (b) a larger scale low-T STM image and the corresponding cross-section along the indicated MTB is shown that measured the periodicity of the CDW as $\sim$ 1.0 nm. The schematic in (c) illustrates the relationship between CDW period and nesting vector $q = 2k_{\rm F}$. Also the opening of a band gap at $k_{\rm F}$ is illustrated. Temperature dependent resistance measurements, shown in (d), indicate two CDW transitions. The transitions at 235 K and 205 K correspond to incommensurate and commensurate CDW transitions, respectively. Depending on the applied bias voltage we also observe a drop in resistance below the CDW transition temperatures, which is attributed to CDW-sliding. The inset shows the control measurement on a bare MoS$_2$ substrate and shows no transitions.}
\label{figure2}
\end{figure}

The mirror twin boundaries in a monolayer transition metal dichalcogenide
\cite{Najmaei-13,Zande-13} are true 1D line defects. They are robust to high temperatures and atmospheric conditions, 
thus making them a promising material system, which is amendable beyond ultra high vacuum investigations and 
useful for potential device fabrication. Previously, the structural properties of these line defects have been studied by
(scanning) transmission electron microscopy\cite{Lehtinen-15,Lin-15,Zande-13,Najmaei-13} and by scanning tunnelling microscopy (STM)
and tunnelling spectroscopy\cite{Liu-15,Liu-14,Barja-16}.\\ \\    
{\bf Results}\\
{\bf Line defect characterisation.} Fig. \ref{figure1} shows STM results of the mono- to bilayer MoSe$_2$ grown on a MoS$_2$ single crystal substrate. Three equivalent directions for the MTBs are observed in the hexagonal MoSe$_2$ crystal. The high density of
these aligned line defects in MoSe$_2$\cite{Zande-13} provides a measurable ARPES signal for this 1DES
and thus enables the $\omega (k)$ characterisation of this line defect.\\ \\
{\bf Peierls transition in MoSe$_2$ grain boundary.} For metallic 1D structures, an instability to 
charge density wave (CDW) is expected (see additional discussion in the supplementary Note 1),
which has been previously reported for MoSe$_2$ grain boundaries by low temperature STM studies \cite{Barja-16}.
The CDW in MTBs gives rise to a tripling of the periodicity, as can be seen in the low temperature-STM 
images shown in Fig. \ref{figure2} (a) and (b). The CDW in 1D metals is a consequence of electron-phonon coupling. 
The real-space periodicity of the CDW is directly related to a nesting of the Fermi wavevector, as schematically 
shown in Fig. \ref{figure2} (c). ARPES measurements of the Fermi-surface can thus directly provide justification for the 
periodicity measured in STM, which is shown below. In addition, the CDW transition is a metal-insulator transition 
and thus changes in the sample resistance occur at the CDW transition temperature. Fig. \ref{figure2} (d) shows a 
4-point measurement with macroscopic contacts on a continuous mono- to bi-layer film (as shown in Fig. \ref{figure1} (c)). Clear jumps 
in the resistance are observed for three different samples at $\sim$ 235 K and $\sim$ 205 K, which 
are attributed to an incommensurate and commensurate CDW transitions, respectively. The drop in resistance 
at lower $T$ is assigned to a depinning of the CDW from defects and so-called CDW sliding. CDW 
sliding is a consequence of the applied potential rather than a specific temperature.

In order to study a stable, gapless, 1DES, we determine the spectral weight together with the energy dispersion in momentum space, 
by performing ARPES measurements at room temperature, which is well above the CDW transition temperature. This 
is done on samples consisting predominantly of monolayer MoSe$_2$ islands, as shown in the supplementary Figure 1.
Fig. \ref{figure3} and supplementary Figure 2 illustrate the Fermi surface of 1D metals, consisting of two parallel lines, 
separated by $2k_{\rm F}$, in the absence of interchain hopping. 
Because of the three equivalent real space directions of the MTBs in our sample, super-positioning of three rotated 1DES
results in star-shaped constant energy surface in reciprocal space, as shown in Fig. \ref{figure3} and supplementary Note 1. In the three cases, 
a perfect nesting is noticeable, namely one complete Fermi sheet can be translated onto the other by a single wave vector  ${\pm}$$2k_{\rm F}$.

Even more important, by using high energy and momentum resolution ARPES, the Fermi-wave vector could be precisely determined, 
giving a value of  $k_{\rm F}= 0.30 \pm 0.02\,\textrm{\AA}^{-1}$, which is about $1/3$ of the BZ-boundary at $\pi/a_{{\rm MoSe}_2}$. Hence 
a band filling of ${n=2/3}$ has been experimentally obtained. The Fermi-wavevector also gives a direct prediction of the CDW periodicity 
of $\pi/k_{\rm F} = 10.5 \pm 0.7\,\textrm{\AA}$, which is in good agreement with $3\times a_{{\rm MoSe}_2}$ measured in 
STM (see Fig. \ref{figure2}).\\ \\
{\bf Spin charge separation.} 
While the perfect nesting conditions in 1D metals predicts a CDW transition, its occurrence is no proof for 1D electron dynamics. 
For obtaining evidence of 1D electron dynamics, a detailed analysis of the spectral function and its consistency with theoretically
predicted dispersions need to be demonstrated. The photoemission spectral function of the 1D state is shown in Figs. \ref{figure3} (e) and (f). 
Without any sophisticated analysis and considering only the raw ARPES data, it is evident that the experimental results are
in complete disagreement with the single dispersing band predicted 
by ground state DFT simulations\cite{Lehtinen-15,Barja-16}. Effectively, our data cannot be fit with a single dispersion branch
(see also supplementary Figure 4 and the supplementary Note 2 for an analysis of the raw data in terms of energy distribution curves (EDC), 
momentum distribution curves (MDC) and lifetime.) 

Using data analysis that applies a curvature procedure to raw data\cite{Zhang-11}, as commonly used in ARPES, the experimental 
band dispersions in the full energy versus momentum space show two clear bands that exhibit  
quite different dispersions. We provisionally associate, which our theoretical results confirms below, the upper and lower dispersion with the spinon
and the holon branch, respectively. Manifestly, the spin mode follows the low-energy part of the 1D parabola, 
whereas the charge mode propagates faster than the spin mode.  The extracted experimental velocity values 
are $v_{\rm h}$= 4.96 $\times$ 10$^{5}$ms$^{-1}$ and $v_{\rm s}$ = 4.37 $\times$ 10$^{5}$ms$^{-1}$, 
revealing a ratio $v_{\rm h}/v_{\rm s}$ of the order of $\approx 0.88$. Notice that these states lie entirely within the band gap of the MoSe$_2$ 
monolayer whose VBM is located at $1.0$ eV below the Fermi-level, see supplementary Figure 3. 

DFT simulations cannot predict the electron removal spectrum of the 1D electron dynamics. Thus the single 
dispersing band obtained in previous DFT simulations for this system is not expected to be consistent with the 
experiment. However, the single-band DFT results indicate that the electron dynamics behaviour can be suitably 
described by a single band Hubbard model and associated PDT. The PDT is a method that has been originally 
used to derive the spectral function of the 1D Hubbard model in the vicinity of high-energy branch-line singularities
\cite{Carmelo-05,Carmelo-06,Carmelo-08,Carmelo-17}.
It converges with TLL for low energies\cite{Carmelo-06B}. As reported below, here we use a renormalized PDT (RPDT) because 
the conventional 1D Hubbard does not include finite-range interactions.\\ \\
{\bf Low energy properties and TLL electron interaction strength.}
Critical for calculating the spectral functions with RPDT is the knowledge 
of the electron interaction strength, which needs to be determined experimentally. Since very close to the Fermi level, 
in the low-energy excitations limit, the RPDT converges to the TLL theory, we have evaluated 
the photoemission weight in the vicinity of the Fermi-level in accordance to TLL theory.
A decisive low-energy property of 1D metals is, according to that theory\cite{Voit-95,Blumenstein-11}, the suppression of the 
DOS at the Fermi-level, whose power law exponent is dependent on the electron interaction range and strength.
Fig. \ref{figure4} shows the angle integrated photoemission intensity, which is proportional to 
the occupied DOS, as a function of energy for the 1DES. It is compared to the photoemission from a gold sample under the same 
conditions. The suppression of the DOS for the 1D defects compared to Au is apparent in Fig. \ref{figure4} (a). 

According to the TLL scheme, the suppression of DOS follows a power law dependence whose exponent is determined by the electron 
interaction strength and range in the 1D system. An exponent of $\sim$ 0.8 is extracted from a log-plot shown in Fig. \ref{figure4} (b). A refined 
fitting for the exponent $\alpha$ that takes the temperature into account\cite{Schonhammer-93} reveals that the data 
are best reproduced for $\alpha$ between $0.75$ and $0.80$ (Fig. \ref{figure4} (c)). The charge TLL parameter $K_{\rm c}$, 
which provides information on the range of the electron interaction\cite{Schulz-90}, is related to $\alpha$ by 
$\alpha = (1-K_{\rm c})^2/4K_{\rm c}$. Hence $K_{\rm c}$ has values between $0.20$ and $0.21$. \\ \\
{\bf Comparison of experiment to the theoretical model.}
Within the 1D Hubbard model with on-site repulsion $U$ and hopping integral $t$, the charge TLL parameter $K_{\rm c}$
and related exponent $\alpha$ values should belong to the ranges $K_{\rm c}\in [1/2,1]$ and $\alpha\in [0,1/8]$, respectively.  
However, our experimental values are in the ranges $K_{\rm c}\in [0.20,0.21]$ and $\alpha\in [0.75,0.80]$, which is an unmistakable signature of 
electron finite-range interactions and therefore our system cannot be studied in the context of the conventional 1D Hubbard model\cite{Schulz-90}. 
Consequently, we have developed a new theoretical scheme that successfully includes such interactions. 
As justified below in the Methods section, the corresponding RPDT specifically relies on the spectral function near the branch lines of the non-integrable 
1D Hubbard model with finite-range interactions being obtained from that of the integrable 1D Hubbard model
PDT\cite{Carmelo-05,Carmelo-06,Carmelo-08,Carmelo-17} upon suitably renormalising its spectra and phase shifts.

The renormalisation using the PDT approach has two steps. The first refers to the $U$ value, which loses its onsite-only character and 
is obtained upon matching the  +experimental band spectra with those obtained within the 1D Hubbard 
model for $n=2/3$, leading to $U = 0.8\,t$. Indeed, the ratio $W_{\rm h}/W_s$
of the observed ${\rm c}$ band (holon) and ${\rm s}$ band (spinon) energy bandwidths $W_{\rm h}=\varepsilon_{\rm c} (2k_{\rm F}) - \varepsilon_{\rm c} (0)$ and 
$W_s =\varepsilon_{\rm s} (k_{\rm F}) - \varepsilon_{\rm s} (0)$, respectively, is achieved for that model at $U/t=0.8$. (The energy dispersions 
$\varepsilon_{\rm c} (q)$ for $q\in [-\pi,\pi]$ and $\varepsilon_{\rm s} (q')$ for $q'\in [-k_{\rm F},k_{\rm F}]$ and the related
$\gamma = {\rm c,c',s}$ exponents ${\tilde{\zeta}}_{\gamma} (k)$ considered in the following are defined in more 
detail in the Methods section.) This renormalisation 
fixes the effective $U$ value yet does not affect $t$. The corresponding ${\rm c}$ and ${\rm c'}$ (holon) and ${\rm s}$ (spinon) branch
lines spectra $\omega_{\rm c} (k) = \varepsilon_{\rm c} (\vert k\vert + k_{\rm F})$ for $k\in [-k_{\rm F},k_{\rm F}]$, $\omega_{\rm c'} (k) = \varepsilon_{\rm c} (\vert k\vert - k_{\rm F})$ for $ \in [-3k_{\rm F},3k_{\rm F}]$ and $\omega_{\rm s} (k) = \varepsilon_{\rm s} (k)$ for $k\in [-k_{\rm F},k_{\rm F}]$ are plotted in 
Figs. \ref{figure5} (d)-(f) and supplementary Figure 5. An important difference relative
to the $n=1$ Mott-Hubbard insulating phase is that for the present $n=2/3$ metallic phase the energy bandwidth 
$W_{\rm c} = \varepsilon_{\rm c} (\pi) - \varepsilon_{\rm c} (2k_{\rm F})$ does not vanish. That the renormalisation does not affect $t$ stems from a symmetry 
that implies that the full ${\rm c}$ band energy bandwidth is independent of both $U$ and $n$ and reads $W_{\rm h} + W_{\rm c} = 4t$. Hence 
$W_{\rm h}=4t$ for the Mott-Hubbard insulator whereas $W_{\rm h}<4t$ for the metal. Combining both the value of the ratio $W_{\rm h}/W_{\rm c}$
for the 1D  Hubbard model at $U/t=0.8$ and $n=2/3$ and the exact relation $W_{\rm h} + W_{\rm c} = 4t$ with analysis of 
Fig. \ref{figure5} (d)-(f), one uniquely finds $t \approx 0.58$ eV. 
The parameter $\alpha$ is here denoted by $\alpha_0$ for the 1D Hubbard model. It reads 
$\alpha_0=(2-\xi_{\rm c}^2)/(8\xi_{\rm c}^2)\in [0,1/8]$ with $\alpha_0 =0$ for $U/t\rightarrow 0$ and $\alpha_0 =1/8$ for $U/t\rightarrow\infty$
where $\xi_{\rm c} = \sqrt{2K_{\rm c}}$ is a superposition of pseudofermion phase shifts. (See Methods.)

The second step of the renormalisation
corresponds to changing the $\xi_{\rm c}$ and phase shift values so that the parameter
$\alpha = (2-{\tilde{\xi}}_{\rm c}^2)^2/(8{\tilde{\xi}}_{\rm c}^2)$ has values in the range $\alpha \in [\alpha_0,\alpha_{\rm max}]$
where $\alpha_0 \approx 1.4\times 10^{-3}$ for $U/t=0.8$ and $n=2/3$. As justified in the Methods section, 
$\alpha_{\rm max}=49/32\approx 1.53$. The effect of increasing $\alpha$ at fixed finite $U/t$ and $n$ from $\alpha_0$ to $1/8$ 
is qualitatively different from that of further increasing it to $\alpha_{\rm max}$. As
discussed in that section, the changes in the $(k,\omega)$ plane weight distribution resulting
from increasing $\alpha$ within the latter interval $\alpha \in [1/8,\alpha_{\rm max}]$ 
are mainly controlled by the finite-range interactions.

For $U/t=0.8$, $n=2/3$ and $T=0$ the one-electron spectral function of both the conventional 
1D Hubbard model ($\alpha = \alpha_0$) and corresponding model with finite range interactions ($\alpha \in [\alpha_0,\alpha_{\rm max}]$)
consists of a $(k,\omega)$-plane continuum within which well-defined singular branch lines emerge. 
Most of the spectral weight is located at and near such singular lines. Near them, the spectral function 
has a power-law behaviour characterised by negative $k$ dependent exponents. At $T\approx 300$\,K such singular lines survive
as features displaying cusps. Our general renormalisation procedure leads to a one-electron 
spectral function expression that for small deviations $(\omega_{\gamma} (k)-\omega)>0$ 
from the finite-energy spectra $\omega_{\gamma} (k)$ of the $\gamma = {\rm c,c',s}$ branch lines
plotted in Fig. \ref{figure5} (d), (e) and (f) reads,
$B (k,\omega) \propto (\omega_{\gamma} (k)-\omega)^{{\tilde{\zeta}}_{\gamma} (k)}$ for
$\alpha \in [\alpha_0,\alpha_{\rm max}]$. 
The singular branch lines correspond to the $\gamma = {\rm c,c',s}$ lines $k$ ranges for which their exponents ${\tilde{\zeta}}_{\gamma} (k)$ are negative. 
As confirmed and justified in the Methods section, for $U/t=0.8$, $n=2/3$ and $t=0.58$\,eV there is quantitative agreement 
with the $(k,\omega)$-plane ranges of the experimentally observed spectral function cusps for $\alpha \in [0.75,0.78]$. 
This is fully consistent with the $\alpha$ experimental uncertainty range $\alpha \in [0.75,0.80]$.
The three $\gamma = {\rm c,c',s}$ exponents momentum dependence for both the 1D Hubbard 
model with finite-range interactions corresponding to $\alpha =0.78$ (full lines) and the conventional 1D Hubbard model
for which $\alpha=\alpha_0\approx 1.4\times 10^{-3}$ (dashed-dotted lines) is plotted in Figs. \ref{figure5} (a), (b) and (c).\\ \\
{\bf Discussion}\\
The astonishing agreement of the theoretical calculations with finite range interactions over the entire 
$(k,\omega)$-plane provides strong evidence for the assignment of the two spectral branches observed in the 
experiments to spin charge separation in a 1D metal. Despite this agreement, alternative explanations for the 
photoemission spectrum should be noted. Strongly asymmetric line shapes in photoemission spectra have been 
reported and thus an assignment of the cusps to yet unknown line-shape effects in 1D materials cannot be entirely 
excluded. However, the accurate prediction of the continuum between the cusp lines and the fit of the 
${\rm c}$ and ${\rm s}$ branch-line dispersions by the 1D Hubbard model with finite range interactions makes alternative 
effects unlikely to reproduce exactly such spectral features. 

Concerning the DOS at the Fermi level, our measurements clearly show a suppression of the DOS that can be fit with a power law behaviour. DOS suppression has, however, also been observed due to final-state pseudogap effects in nanostructures\cite{Joynt-99,Hovel-98}. While it is difficult to 
exclude such effects categorically, the expected 1D nature of the line defects and thus the breakdown of Fermi-liquid 
theory requires application of TLL, as has been applied to other (quasi) 1D systems in the past 
\cite{Ohtsubo-15,Blumenstein-11,Ishii-03}, to interpret photoemission intensity at the Fermi level. 
Certainly, obtaining the same exponent  $\alpha$ for the power law behaviour of TLL from the experimental fit of the 
DOS and the spectral features of the 1D Hubbard model with finite range interactions support the assignment of 
the DOS suppression at the Fermi-level to TLL effects. 

We have presented a detailed experimental analysis of the electronic structure of a material 
line defect by angle resolved photoemission. High density of twin grain boundaries in epitaxial monolayer 
MoSe$_2$ could be analyzed by angle resolved photoemission spectroscopy. This enabled us to accurately determine 
the Fermi surface and demonstrate the CDW observed in this material is a consequence of Fermi wave vector nesting. 
Both the suppression of DOS at the Fermi level as well as broad spectral features with notable cusps are in agreement 
with 1D electron dynamics. 
While the low-energy spectra are described by TLL, the dispersion of the cusps in the full energy versus momentum space 
in high-energy range could be only
accurately reproduced by a 1D Hubbard model with suitable finite range interactions. Consequently, the cusps could  be interpreted 
as spin- and charge- separation in these 1D metals. The accurate description of the experiment by RPDT calculations allows 
us to go beyond the low energy restriction of TLL, showing that the exotic 1D physics is valid for both low- and high-energy, 
with non-linear band dispersions and broad momentum values. Unlike other systems that only exhibit strong 1D anisotropy, 
the intrinsic line defects in TMDs have no specific repetition length and can thus be viewed as true 1D structures. Moreover, 
isolated twin grain boundaries of micrometer length have been recently reported in CVD-grown TMDs\cite{Najmaei-13}, 
which can be envisaged  as remarkable candidates for quantum transport measurements on isolated 1D metals. Furthermore, 
2D materials can be gated and this will exert control of transport properties of these quantum wires.\\  \\
{\bf Methods}\\ 
{\bf Sample preparation.} Monolayer MoSe$_2$ islands were grown by van der Waals epitaxy by co-deposition of atomic Se 
from a hot wall Se-cracker source and Mo from a mini-e-beam evaporator. The MoS$_2$ single crystal substrate was a 
synthetically grown and cleaved in air before introducing into the UHV chamber where it was outgassed at 300$^{\circ}$C for 4 
hours prior to MoSe$_2$ growth. Mo has been deposited in a selenium rich environment at a substrate temperature 
of $\sim$ 300-350$^{\circ}$C. The MoSe$_2$ monolayer was grown slowly with a growth rate of $\sim$ 0.16 monolayers per hour. 
While the detailed mechanism for the formation of MTBs during MBE growth is not completely understood, it has been noted that 
the structure shown in Fig. \ref{figure1} (a) is deficient in chalcogen atoms, i.e. the grain boundary has a stoichiometry of MoSe 
embedded in the MoSe$_2$ matrix. Computational studies have shown that MTBs are thermodynamically favoured over 
the formation of high density of individual chalcogen vacancies\cite{Lehtinen-15} and this may explain their presence in MBE grown samples. 
These samples were investigated by RT STM in a surface analysis chamber connected to the growth chamber. Additional 
characterisation by VT-STM and ARPES were performed by transferring the grown samples in a vacuum suitcase to 
the appropriate characterisation chambers. In addition, air-exposed samples were characterised by ARPES. After vacuum 
annealing to $\sim$ 300$^{\circ}$C, the ARPES results were indistinguishable to the in vacuum transferred samples indicating the 
stability of the material in air against oxidation and other degradation. The stability of the sample also enables the four-point 
transport measurements described below.\\ \\ 
{\bf ARPES measurements.} Micro-ARPES measurements were performed at the ANTARES beamline at the SOLEIL 
synchrotron. The beam spot size was $\sim$ 120 $\mu$m. The angular and energy resolution of the beamline at a photon 
energy of 40 eV are $\sim$ 0.2$^{\circ}$ and $\sim$ 10 meV, respectively. Most of the data were collected around the $\Gamma$-point 
of the second Brillouin zone, corresponding to an emission angle of 42.5$^{\circ}$ degrees with respect to the surface normal, for photon 
energy of 40 eV. Both left and right circular polarized light as well as linear polarized light was used. The photon-incident 
angle on the sample was normal incidence. For circular polarized light photoemission from all MTBs is obtained. Emission 
from a single MTB direction could be enhanced with linear polarized light and the {\bf A}-vector parallel to the surface. For azimuth 
rotation with the {\bf A}-vector aligned to the direction of one MTB enhanced emission from this direction was obtained as shown 
in Fig. \ref{figure3} (c). All data shown here were obtained at 300 K.\\ \\
{\bf Broadening of the ARPES spectral function and lifetime analysis.} As it has already been reported in previous ARPES studies (see for instance Fig. 5 of Ref. \cite{Kim-06}), the lifetime of a Fermi-liquid quasi-particle, $\tau(k)$, can be directly determined from the width of the peak in the energy distribution curves (EDC), analysing the ARPES 
data defined by the spectral weight at fixed $k$ as a function of $\omega $, where $\omega $ is the energy. Specifically, 
\begin{equation}
1/\tau(k) = \Delta\omega \, .
\label{tau}
\end{equation}
The consistency of a Fermi-liquid picture can be also checked by studying the momentum distribution curves (MDC), i.e., from the momentum width $\Delta k$ of the spectral function 
peak at fixed binding energy, $\omega $. As long as the Fermi-liquid quasi-particle excitation is well defined, (i.e., the decay rate is small 
compared to the binding energy), the energy bandwidth and momentum width are related as, 
\begin{equation}
\Delta\omega = v_{\rm F}\,\Delta k \, .
\label{Domega}
\end{equation}
Here $v_{\rm F}$ is the renormalized Fermi velocity, which can be directly measured using high energy and momentum resolution ARPES. 
Due to the separation of charge and spin, one hole (or one electron) is always unstable to decay into two or more elementary excitations, 
of which one or more carries its spin and one or more carries its charge. 
Then elementary kinematics implies that, at T= 0, the spectral function is nonzero only for negative frequencies such that, 
\begin{equation}
\vert\omega\vert  \leq {\rm{min}} (v_{\rm c},v_{\rm s})\vert k \vert \, ,
\label{minvcvs}
\end{equation}
where $v_{\rm c}$ and $v_{\rm s}$ are the charge and spin velocity, respectively. This analysis procedure is described in Figure 6, where the spectral function particularly at $\omega $ values between $0.40$ eV up to $0.95$ eV shows a continuum, which is valid 
for all momentum $k$ values that fit Eq. (\ref{minvcvs}). MDC and EDC plots are sensitive to this detachment of the system with respect 
to a conventional Fermi-liquid quasi-particle behaviour. 

This type of analysis, based on the shape of EDC and MDC plots, is also well explained by 
Emery et al. (see Fig. 2 and Fig. 3 of Ref. [\cite{Orgad-01}]). In Fig. \ref{figure6} we  present the results of a similar analysis. As it is shown 
in panels (d) and (e), the MDC and EDC cuts of the raw data at different binding energies and momentum, respectively, show a clear 
enlargement of the lifetime that can be extracted from the ARPES data. However, this experimental value is just proportional to 
various interaction strengths. This approximative methodology of the nature and magnitude of the present interactions can be improved by using more sophisticate 
theoretical approaches as the one reported in the present manuscript.\\ \\
{\bf PDT as starting point of our theoretical method.}
The method used in our theoretical analysis of the spin-charge separation observed in the 1D quantum-line defects of
MoSe$_2$ was conceived for that specific goal. It combines the pseudofermion dynamical theory (PDT) for the 1D Hubbard model\cite{Carmelo-05,Carmelo-17,Carmelo-06B}
with a suitable renormalisation procedure.

On the one hand, the 1D Hubbard model range $\alpha_0 \in [0,1/8]$ corresponds to the intervals $K_{\rm c} \in [1/2,1]$ and $\xi_{\rm c} \in [1,\sqrt{2}]$
of the TLL charge parameter\cite{Schulz-90,Voit-95,Pham-00} and the related parameter $\xi_{\rm c} = \sqrt{2K_{\rm c}}$.
On the other hand, the range $\alpha\in [0.75,0.78]$ for which the renormalised theory is found 
to agree with the experiments implies that ${\tilde{K}}_{\rm c} =1 + 2\alpha - 2\sqrt{\alpha (1+\alpha)}$
and ${\tilde{\xi}}_{\rm c} = \sqrt{2{\tilde{K}}_{\rm c}}$ have values in the ranges ${\tilde{K}}_{\rm c} \in [0.20,0.21]$ 
and ${\tilde{\xi}}_{\rm c} \in [0.63,0.65]$, respectively. Here ${\tilde{K}}_{\rm c} $
and ${\tilde{\xi}}_{\rm c}$ is our notation for the TTL charge parameter and related parameter, respectively,
in the general case when they may have values within the extended intervals 
${\tilde{K}}_{\rm c} \in [1/8,1]$ and thus ${\tilde{\xi}}_{\rm c} \in [1/2,\sqrt{2}]$. The minimum values 
${\tilde{K}}_{\rm c} =1/8$ and ${\tilde{\xi}}_{\rm c} = 1/2$ follow from corresponding phase-shift allowed ranges. (Below the relation
of ${\tilde{\xi}}_{\rm c}$ to phase shifts is reported.) The above experimental subinterval ${\tilde{K}}_{\rm c} \in [0.20,0.21]$
belongs to the interval ${\tilde{K}}_{\rm c} \in [1/8,1/2]$ for which the electron finite-range interactions must be accounted for\cite{Schulz-90}.

In the case of the conventional 1D Hubbard model, the PDT was the first approach to compute the spectral functions 
for finite values of $U/t$ near singular lines at high energy scales beyond the low-energy 
TLL limit\cite{Carmelo-05}. (In the low-energy limit the PDT recovers
the TLL physics\cite{Carmelo-06B}.) After the PDT was introduced for that integrable model, novel methods that rely on a 
mobile impurity model (MIM) approach have been developed to tackle the high-energy physics of both non-integrable and
integrable 1D correlated quantum problems, also beyond the low-energy TLL limit
\cite{Glazman-09,Glazman-12,Essler-10,Seabra-14}. The relation between the PDT and MIM has been
clarified for a simpler model\cite{Carmelo-16}, both schemes leading to exactly the same momentum 
dependent exponents in the spectral functions expressions. Such a relation applies as well to more complex models.
For instance, studies of the 1D Hubbard model by means of the MIM\cite{Essler-10,Seabra-14} lead to exactly the 
same momentum, interaction, and density dependence as the PDT for the exponents that control the 
one-electron removal spectral function near its branch lines. 

For integrable models, in our case the 1D Hubbard model, there is a representation in terms of elementary objects called
within the PDT ${\rm c}$ and ${\rm s}$ pseudofermions for which there is only zero-momentum forward-scattering at all energy
scales. The ${\rm c}$ and ${\rm s}$ bands momentum values are associated with the 1D Hubbard model exact Bethe-ansatz solution quantum
numbers. The ${\rm c}$ pseudofermion and the ${\rm s}$ pseudofermion annihilated under transitions from the $N$ electron ground state to the 
$N-1$ electron excited states refer to the usual holon and spinon, respectively\cite{Voit-95,Pham-00,Glazman-12}. 

That for the pseudofermions there is only zero-momentum 
forward-scattering at all energy scales, follows from the existence of an infinite number of conservation laws associated
with the model integrability\cite{Shastry-86,Shastry-86B}. This means that in contrast to the model 
underlying electron interactions, the pseudofermions,
upon scattering off each other only acquire phase shifts. Hence under their scattering events there is no energy 
and no momentum exchange, on the contrary of the more complex underlying physical particles interactions.
In the vicinity of well-defined $(k,\omega)$-plane features called branch lines, the $T=0$ spectral functions
of integrable 1D correlated models are of power-law form with negative momentum dependent exponents. 
Such properties apply to all integrable 1D correlated models.\\ \\ 
{\bf Universality behind our method renormalisation procedures.}
In the case of non-integrable 1D correlated models, there is no pseudofermion representation for which there is only zero-momentum 
forward-scattering at all energy scales. This is due to the lack of an infinite number of conservation laws.
The universality found in the framework of the MIM for the spectral functions of non-integrable and integrable 1D models 
\cite{Glazman-09,Glazman-12} refers to specific energy scales corresponding to both the low-energy TLL spectral features 
and energy windows near the high-energy non-TLL branch lines singularities. In the vicinity of these lines, the $T=0$ spectral functions
of non-integrable 1D correlated models are also of power-law form with negative momentum dependent exponents. 

This universality means that at both these energy scales there is for such models a suitable representation in terms of 
pseudofermions that undergo only zero-momentum forward-scattering events and whose phase shifts control the spectral 
functions behaviours. Our renormalisation scheme for adding electron finite-range interactions to the 1D Hubbard model and
corresponding PDT relies on this universality. Indeed, the finite-range interactions render the model non-integrable. 
However, in the vicinity of the branch lines singularities the spectral function remains having the same universal behaviour. 
Our normalisation procedure can be used for any chosen $\alpha$ value in the range $\alpha \in [\alpha_0,\alpha_{\rm max}]$. 
Here $\alpha_0\in [0,1/8]$ is the conventional 1D Hubbard model $\alpha$ value for given $U/t$ and electronic density $n$ values. 
For the $U/t=0.8$ and $n=2/3$ values found within our description of the 1D quantum-line defects of MoSe$_2$ it reads 
$\alpha_0\approx 1.4\times 10^{-3}$. The maximum $\alpha$ value $\alpha_{\rm max}=49/32=1.53125$ 
refers through the relation $\alpha = (1-{\tilde{K}}_{\rm c})^2/4{\tilde{K}}_{\rm c}$ and thus $\alpha = (2-{\tilde{\xi}}_{\rm c}^2)^2/8{\tilde{\xi}}_{\rm c}^2$
to the above minimum values ${\tilde{K}}_{\rm c}=1/8$ and ${\tilde{\xi}}_{\rm c} =1/2$. 

The renormalisation of the conventional 1D Hubbard model used in our studies refers to some 1D Hamiltonian with the same terms as 
that model plus finite-range interaction terms. The latter terms are neither a mere first-neighbouring $V$ term nor
a complete long-range Coulomb potential extending over all lattice sites. Interestingly, the specific
form of the additional finite-range interaction Hamiltonian terms is not needed for our study. This follows from
the above universality implying that both for the low-energy TLL limit and energy windows near the high-energy branch lines singularities
of the 1D Hubbard model with finite-range interactions under consideration the relation of $\alpha$ to the phase 
shifts remains exactly the same as for the conventional 1D Hubbard model.

Importantly, the only input parameters of our renormalisation procedure
are the effective $U$ and transfer integral $t$ values for which the theoretical branch 
lines energy bandwidths match the corresponding experimental bandwidths. Apart from the 
1D quantum-line defects band-filling $n=2/3$, our approach has no additional ``fitting parameters''.\\ \\ 
{\bf The spectra in terms of pseudofermion energy dispersions.} 
Within the PDT for the 1D Hubbard model\cite{Carmelo-05,Carmelo-06,Carmelo-08,Carmelo-17},
nearly the whole electron removal spectral weight is in the metallic phase
originated by two $\iota=\pm 1$ excitations generated from the ground 
state by removal of one ${\rm c}$ pseudofermion of momentum $q\in [-2k_{\rm F},2k_{\rm F}]$ and one ${\rm s}$ pseudofermion of momentum 
$q'\in [-k_{\rm F},k_{\rm F}]$. The superposition in the $(k,\omega)$-plane of the spectral weights associated
with the corresponding two $\iota = \pm 1$ spectra generates the multi-particle continuum. Such $\iota = \pm 1$
spectra are of the form,
\begin{eqnarray}
\omega (k) & = & \varepsilon_{\rm c} (q) + \varepsilon_{\rm s} (q')\leq 0
\nonumber \\
k & = & -\iota\,2k_{\rm F} - q - q' \, , \hspace{0.5cm} \iota = \pm 1 \, .
\label{2paspec}
\end{eqnarray}
They are two-parametric, as they depend on the two independent ${\rm c}$ and ${\rm s}$ bands momenta $q$ and $q'$,
respectively. Hence such spectra refer to two-dimensional domains in the $(k,\omega)$-plane. 
They involve the energy dispersion $\varepsilon_{\rm c} (q)$ whose ${\rm c}$ momentum band
interval is $q\in [-\pi,\pi]$ and whose ground-state ${\rm c}$ pseudofermion occupancy is $q\in [-2k_{\rm F},2k_{\rm F}]$ and 
the dispersion $\varepsilon_{\rm s} (q')$ whose ${\rm s}$ momentum band range is $q'\in [-k_{\rm F},k_{\rm F}]$, which is full
in the present zero spin-density ground state, are defined below. 

The multi-particle continuum in the one-electron removal spectral function that results from the superposition of 
the spectral weights associated with the two $\iota =\pm 1$ spectra contains three branch lines that display the cusps:
two $c,\iota$ branch lines and a ${\rm s}$ branch line. The $c,\iota$ branch lines result from processes for which 
the removed ${\rm c}$ pseudofermion has momentum in the range $q\in [-2k_{\rm F},2k_{\rm F}]$ and the 
removed ${\rm s}$ pseudofermion has momentum $q' = -\iota k_{\rm F}=\mp k_{\rm F}$. Hence the excitation
physical momentum is $k =-\iota\,k_{\rm F} - q=\mp k_{\rm F} - q$. The ${\rm s}$ branch line results from removal of one
${\rm c}$ pseudofermion of momentum $q=-\iota\,2k_{\rm F}=\mp 2k_{\rm F}$. The removed ${\rm s}$ pseudofermion has
momentum in the interval $q'\in [-k_{\rm F},k_{\rm F}]$. The physical momentum is then given by $k=-q'$.

It is convenient to redefine the two $c,\iota$ branch lines in terms of related ${\rm c}$ and ${\rm c'}$ branch lines. The
spectra of the ${\rm c}$, ${\rm c'}$, and ${\rm s}$ branch lines are plotted in Figs. 5 (d), (e) and (f) for $U/t=0.8$,
$t =0.58$ eV and electronic density $n=2/3$. On the one hand, the ${\rm c}$ branch line results from processes relative to 
the ground state that involve removal of one ${\rm c}$ pseudofermion with momentum belonging to the ranges $q \in [-2k_{\rm F},-k_{\rm F}]$ 
and $q \in [k_{\rm F},2k_{\rm F}]$ and removal of one ${\rm s}$ pseudofermion with momentum $q' = - \iota\,k_{\rm F}$ for $\iota =  {\rm sgn}\{k\}$. 
The ${\rm c}$ branch line spectrum then reads,
\begin{eqnarray}
\omega_{\rm c} (k) & = & \varepsilon_{\rm c} (\vert k\vert + k_{\rm F}) 
\nonumber \\
k & = & -{\rm sgn}\{k\} k_{\rm F} - q \in [-k_{\rm F},k_{\rm F}] \, .
\label{omegac}
\end{eqnarray}
On the other hand, the ${\rm c'}$ branch line is generated by removal of one ${\rm c}$ 
pseudofermion with momentum belonging to the ranges $q \in [-2k_{\rm F},k_{\rm F}]$ and $q \in [-k_{\rm F},2k_{\rm F}]$ and removal of one
${\rm s}$ pseudofermion with momentum $q' = - \iota\,k_{\rm F}$ for $\iota = -{\rm sgn}\{k\}$. Its spectrum is given by,
\begin{eqnarray}
\omega_{\rm c'} (k) & = & \varepsilon_{\rm c} (\vert k\vert - k_{\rm F}) 
\nonumber \\
k & = & {\rm sgn}\{k\} k_{\rm F} - q \in [-3k_{\rm F},3k_{\rm F}] \, .
\label{omegacline}
\end{eqnarray}
The ${\rm s}$ branch line spectrum reads,
\begin{eqnarray}
\omega_{\rm s} (k) & = & \varepsilon_{\rm s} (k) 
\nonumber \\
k & = & -q' \in [-k_{\rm F},k_{\rm F}] \, .
\label{omegas}
\end{eqnarray}

The dispersions $\varepsilon_{\rm c} (q)$ and $\varepsilon_{\rm s} (q')$ appearing in these equations 
are uniquely defined by the following equations valid for $U/t>0$ 
and electronic densities $n\in [0,1]$,
\begin{eqnarray}
\varepsilon_{\rm c} (q) & = & {\bar{\varepsilon}_{\rm c}} (k (q))\hspace{0.35cm}{\rm for}\hspace{0.25cm}q \in [-\pi,\pi] 
\nonumber \\
\varepsilon_{\rm s} (q') & = & {\bar{\varepsilon}_{\rm s}} (\Lambda (q'))\hspace{0.35cm}{\rm for}\hspace{0.25cm}q' \in [-k_{\rm F},k_{\rm F}] \, ,
\nonumber \\
{\bar{\varepsilon}_{\rm c}} (k) & = & \int_Q^{k}dk^{\prime}\,2t\,\eta_{\rm c} (k^{\prime})\hspace{0.35cm}{\rm for}\hspace{0.25cm}k \in [-\pi,\pi]
\nonumber \\
{\bar{\varepsilon}_{\rm s}} (\Lambda) & = & \int_{\infty}^{\Lambda}d\Lambda^{\prime}\,2t\,\eta_{\rm s} (\Lambda^{\prime})
\hspace{0.35cm}{\rm for}\hspace{0.25cm}\Lambda \in [-\infty,\infty] \, . 
\label{varepsilon-s}
\end{eqnarray}
Here the distributions $2t\,\eta_{\rm c} (\Lambda)$ and $2t\,\eta_{\rm s} (\Lambda)$ are the unique solutions of coupled
integral equations given in supplementary Equations (1) and (2). 

The $q$ and $q'$ dependence of the dispersions $\varepsilon_{\rm c} (q)$ and $ \varepsilon_{\rm s} (q')$
occurs through that of the momentum rapidity function $k = k (q)$ for $q \in [-\pi,\pi]$ and
spin rapidity function $\Lambda = \Lambda (q')$ for $q' \in [-k_{\rm F},k_{\rm F}]$, respectively.
Those are defined in terms of their inverse functions $q = q (k)$ for $k \in [-\pi,\pi]$ and
$q' = q' (\Lambda)$ for $\Lambda \in [-\infty,\infty]$ in the supplementary Equations (3) and (4).
The distributions $2\pi\rho (k)$ and $2\pi\sigma (\Lambda)$ in their expressions are the
unique solutions of the coupled integral equations provided in supplementary Equations (5) and (6).\\ \\
{\bf Spectral function within the conventional 1D Hubbard model.} 
Within the PDT for the 1D Hubbard model\cite{Carmelo-05,Carmelo-06,Carmelo-08,Carmelo-17}, the spectral weight distributions are controlled by the
set of phase shifts $\pm 2\pi\Phi_{\beta,\beta'}(q,q')$ acquired by the $\beta = {\rm c}$ and $\beta = {\rm s}$ pseudofermions with momentum $q$ upon 
scattering off each $\beta' = {\rm c}$ and $\beta' = {\rm s}$ pseudofermion with momentum $q'$ created $(+)$ or annihilated $(-)$ under
the transitions from the ground state to the excited energy eigenstates. (In contrast to otherwise in this section, here the
momentum values $q$ and $q'$ are not necessarily those of ${\rm c}$ and ${\rm s}$ pseudofermions, respectively.) 

The expressions of the momentum dependent exponents that control the line shape in the vicinity of the $\gamma = {\rm c, c',s}$ branch lines involve 
phase shifts whose $\beta = {\rm c,s}$ pseudofermions have momentum at the corresponding Fermi points, $\pm q_{{\rm Fc}} = \pm 2k_{\rm F}$ and 
$\pm q_{{\rm Fs}} = \pm k_{\rm F}$. This includes phase shifts $2\pi\Phi_{\beta,\beta'}(\iota q_{{\rm F}\beta},\iota' q_{{\rm F}\beta'})=-2\pi\Phi_{\beta,\beta'}(-\iota q_{{\rm F}\beta},-\iota' q_{{\rm F}\beta'})$, where $\iota =\pm 1$, $\iota' =\pm 1$, acquired by such $\beta = {\rm c,s}$ pseudofermions
upon scattering off $\beta'= {\rm c,s}$ pseudofermions of momentum also at Fermi points annihilated under the transitions from the
$N$ electron ground state to the $N-1$ excited states. Furthermore, such exponents expressions also involve
phase shifts $-2\pi\Phi_{\beta,c}(q_{{\rm F}\beta},q)=2\pi\Phi_{\beta,c}(-q_{{\rm F}\beta},-q)$
and $-2\pi\Phi_{\beta,s}(q_{{\rm F}\beta},q')=2\pi\Phi_{\beta,s}(-q_{{\rm F}\beta},-q')$ acquired by the same $\beta = {\rm c,s}$ pseudofermions
upon scattering off $\beta' ={\rm c}$ and $\beta' ={\rm s}$ pseudofermions of momentum 
$q\in [-2k_{\rm F},2k_{\rm F}]$ and $q'\in [-k_{\rm F},k_{\rm F}]$, respectively, annihilated under such transitions. 

For energy windows corresponding to small energy deviations 
$(\omega_{\gamma} (k)-\omega)>0$ from the high-energy $\gamma = {\rm c,c',s}$ branch-line spectra
$\omega_{\rm c} (k) = \varepsilon_{\rm c} (\vert k\vert + k_{\rm F})$ for $k \in [-k_{\rm F},k_{\rm F}]$,
$\omega_{\rm c'} (k) = \varepsilon_{\rm c} (\vert k\vert - k_{\rm F})$ for $k \in [-3k_{\rm F},3k_{\rm F}]$ and
$\omega_{\rm s} (k) = \varepsilon_{\rm s} (k)$ for $k \in [-k_{\rm F},k_{\rm F}]$, Eqs. (\ref{omegac})-(\ref{omegas}), the electron removal spectral function
has within the PDT the universal form\cite{Carmelo-06,Carmelo-08,Carmelo-17,Carmelo-06B},
\begin{equation}
B (k,\omega) \propto (\omega_{\gamma} (k)-\omega)^{\zeta_{\gamma} (k)}\hspace{0.35cm}{\rm for}\hspace{0.25cm}
\gamma = {\rm c,c',s} \, .
\label{BHM}
\end{equation}
The exponents in this general expression are for $U/t>0$ and electronic densities $n\in [0,1]$ given 
in terms of pseudofermion phase shifts in units of $2\pi$ by,
\begin{eqnarray}
\zeta_{\rm c} (k) & = & -{1\over 2} + \sum_{\iota=\pm1}\left({\xi_{\rm c}\over 4} + {\rm sgn}\{k\}\Phi_{\rm c,c}(\iota 2k_{\rm F},q)\right)^2  
\nonumber \\
k & = & \in [-k_{\rm F},k_{\rm F}] \, , 
\nonumber \\
q & = & -{\rm sgn}\{k\} k_{\rm F} - k \in [-2k_{\rm F},-k_{\rm F}]\,;[k_{\rm F},2k_{\rm F}] \, ,
\nonumber \\
\zeta_{\rm c'} (k) & = & -{1\over 2} + \sum_{\iota=\pm1}\left({\xi_{\rm c}\over 4} - {\rm sgn}\{k\}\Phi_{\rm c,c}(\iota 2k_{\rm F},q)\right)^2  
\nonumber \\
k & = & \in [-3k_{\rm F},3k_{\rm F}] \, , 
\nonumber \\
q & = & {\rm sgn}\{k\} k_{\rm F} - k \in [-2k_{\rm F},k_{\rm F}]\,;[-k_{\rm F},2k_{\rm F}] \, .
\nonumber \\
\zeta_{\rm s} (k) & = & -1 + \sum_{\iota=\pm1}\left({\iota\over 2\xi_{\rm c}} + \Phi_{\rm c,s}(\iota 2k_{\rm F},q')\right)^2  
\nonumber \\
k & \in & [-k_{\rm F},k_{\rm F}]\hspace{0.25cm}{\rm and}\hspace{0.25cm}q' = -k \in  [-k_{\rm F},k_{\rm F}] \, .
\label{3expoH}
\end{eqnarray}

At zero spin density, the entries of the conformal-field theory dressed-charge matrix $Z$
and corresponding matrix $(Z^{-1})^T $ can be alternatively expressed 
in terms of pseudofermion phase shifts in units of $2\pi$ and of the related
parameters $\xi_{\rm c}$ and $\xi_{\rm s}$, as given in supplementary Equations (7) and (8), respectively.
(Here we use the dressed-charge matrix definition of Ref. [\cite{Carmelo-06B}], which is the transposition of that of 
Ref. [\cite{Pham-00}].) Conversely, the pseudofermion phase shifts with both momenta at the Fermi points can be expressed 
in terms of only the  charge TLL parameter $K_{\rm c} = \xi_{\rm c}^2/2$ and 
spin TLL parameter $K_{\rm s} = \xi_s^2/2$\cite{Pham-00} and thus of the present
related $\beta = {\rm c,s}$ parameters $\xi_{\beta} = \sqrt{2K_{\beta}}$. Specifically,
\begin{eqnarray}
& & 2\pi\Phi_{\beta,\beta'}(\iota\,q_{{\rm F}\beta},q_{{\rm F}\beta'}) = \iota\,2\pi\Phi_{\beta,\beta'}(q_{{\rm F}\beta},\iota\,q_{{\rm F}\beta'}) 
\nonumber \\
& & = {\pi\,(\xi_{\beta} -1)^2\over \xi_{\beta}}\hspace{0.35cm}{\rm for}\hspace{0.25cm}\beta = \beta' \, , \hspace{0.5cm} \iota = +1 \, ,
\nonumber \\
& & = - {\pi\,(\xi_{\beta}^2 -1)\over \xi_{\beta}}\hspace{0.35cm}{\rm for}\hspace{0.25cm}\beta = \beta' \, , \hspace{0.5cm} \iota = -1 \, ,
\nonumber \\
& & = (-\iota)^{\delta_{\beta,s}}{\pi\over 2}\,\xi_{\beta}\hspace{0.35cm}{\rm for}\hspace{0.25cm}\beta \neq \beta' \, , \hspace{0.5cm} \iota = \pm 1 \, .
\label{PhiFP}
\end{eqnarray}
Here $\beta = {\rm c,s}$ and $\beta' = {\rm c,s}$. 

The two sets of two coupled integral equations, supplementary Equations (1)-(2) and
(5)-(6), respectively, that one must solve to reach the momentum dependence of the
exponents, Eq. (\ref{3expoH}), have no simple analytical solution. Within our study, these equations are solved
by exact numerical methods. The exponents found from such a numerical solution are plotted as a function of the momentum 
$k$ in Figs. 5 (a), (b) and (c) (dashed-dotted lines) for $U/t=0.8$, $t=0.58$\,eV and electronic density $n=2/3$.
The ${\rm c}$ and ${\rm s}$ exponent expressions in Eq. (\ref{3expoH}) are not valid at the low-energy limiting values $k=\pm k_{\rm F}$.

In the present zero spin-density case, the spin $SU(2)$ symmetry implies that the parameter $\xi_{\rm s}$ appearing 
in Eq. (\ref{PhiFP}) is $u$ independent and reads $\xi_{\rm s} = \sqrt{2}$. The parameter $\xi_{\rm c}$ in 
Eqs. (\ref{3expoH}) and (\ref{PhiFP}) is in turn given by $\xi_{\rm c} = f (\sin Q/u)$ where the function 
$f (r)$ is the unique solution of the integral equation given in the supplementary Equation (9) whose
kernel $D (r)$ is defined in supplementary Equation (10). The parameter $\xi_{\rm c} \in [1,\sqrt{2}]$ has limiting 
values $\xi_{\rm c}=\sqrt{2}$ for $u\rightarrow 0$ and $\xi_{\rm c}=1$ for $u\rightarrow\infty$. 
This is why for the 1D Hubbard model the exponent in the low-$\omega$ power law dependence of the electronic density 
of states suppression $\vert\omega\vert^{\alpha_0}$, 
\begin{equation}
\alpha_0 = {(1-K_{\rm c})^2\over 4K_{\rm c}} = {(2-\xi_{\rm c}^2)^2\over 8\xi_{\rm c}^2} \in [0,1/8] \, ,
\label{alphaHM}
\end{equation}
has corresponding limiting values $\alpha_0=0$ for $u\rightarrow 0$ and $\alpha_0=1/8$ for $u\rightarrow\infty$. 

The ${\rm c}$ pseudofermion phase shifts $2\pi\Phi_{\rm c,c}(\iota 2k_{\rm F},q)$ for $q\in [-2k_{\rm F},2k_{\rm F}]$ and $2\pi\Phi_{\rm c,s}(\iota 2k_{\rm F},q')$ 
for $q'\in [-k_{\rm F},k_{\rm F}]$ that determine the momentum dependence of the exponents in Eq. (\ref{3expoH}) are beyond the reach of the TTL.
Such exponents also involve the ${\rm s}$ pseudofermion phase shifts $2\pi\Phi_{\rm s,c}(\iota k_{\rm F},q)$
and $2\pi\Phi_{\rm s,s}(\iota k_{\rm F},q')$. Due to the spin $SU(2)$ symmetry, at zero spin density the latter phase shifts are $u$ independent.
They are given in the supplementary Equations (14) and (15). Their values provided in these equations
have been accounted for in the derivation of the exponents expressions in Eq. (\ref{3expoH})
and contribute to them. 

The ${\rm c}$ pseudofermion phase shifts explicitly appearing in the exponents expressions, Eq. (\ref{3expoH}), can be written as
$2\pi\Phi_{\rm c,c}\left(\iota 2k_{\rm F},q\right) = 2\pi\bar{\Phi }_{\rm c,c} \left(\iota\sin Q/u,\sin k (q)/u\right)$
and $2\pi\Phi_{\rm c,s}\left(\iota 2k_{\rm F},q'\right) = 2\pi\bar{\Phi }_{\rm c,s} \left(\iota\sin Q/u,\Lambda (q')/u\right)$
where the parameters $\pm Q = k (\pm 2k_{\rm F})$ define the ${\rm c}$ pseudofermion Fermi points in rapidity space.
The corresponding general ${\rm c}$ pseudofermion phase shifts are given by
$2\pi\Phi_{\rm c,c}\left(q,q'\right) = 2\pi\bar{\Phi }_{\rm c,c} \left(\sin k (q)/u,\sin k (q')/u\right)$
and $2\pi\Phi_{\rm c,s}\left(q,q'\right) = 2\pi\bar{\Phi }_{\rm c,s} \left(\sin k (q)/u,\Lambda (q')/u\right)$
where the related rapidity phase shifts $2\pi\bar{\Phi }_{\rm c,c} (r,r')$ and $2\pi\bar{\Phi }_{\rm c,s} (r,r')$
are the unique solutions of the integral equations given in the supplementary Equations (11) and (12). The
free term $D_0 (r)$ of the former integral equation is provided in supplementary Equation (13).

One finds from manipulations of integral equations that the energy dispersions $\varepsilon_{\rm c} (q)$ and $\varepsilon_{\rm s} (q)$, Eq. (\ref{varepsilon-s}),
can be expressed exactly in terms of the ${\rm c}$ pseudofermion rapidity phase shifts as follows,
\begin{eqnarray}
& & \varepsilon_{\rm c} (q) = \varepsilon_{\rm c}^0 (q) - \varepsilon_{\rm c}^0 (2k_{\rm F}) \, ,
\nonumber \\
& & \varepsilon_{\rm c}^0 (q) = -2t\cos k (q) 
\nonumber \\
& & + {t\over \pi}\int_{-Q}^{Q}dk\,2\pi\bar{\Phi }_{\rm c,c}
\left({\sin k\over u}, {\sin k (q)\over u}\right)\sin k  \, ,
\label{epsilon-cq}
\end{eqnarray} 
and 
\begin{eqnarray}
\varepsilon_{\rm s} (q') & = & \varepsilon_{\rm s}^0 (q') - \varepsilon_{\rm s}^0 (k_{\rm F}) \, ,
\nonumber \\
\varepsilon_{\rm s}^0 (q') & = &
{t\over \pi}\int_{-Q}^{Q}dk\,2\pi\bar{\Phi }_{\rm c,s}
\left({\sin k\over u}, {\Lambda (q')\over u}\right)\sin k \, ,
\label{epsilon-sq}
\end{eqnarray} 
respectively. Here $k = k (q)$ and $\Lambda = \Lambda (q')$ are
the momentum rapidity function and spin rapidity function, respectively,
considered above.\\ \\
{\bf Description of the finite-range interactions within our method.}
Below it is confirmed that except for the effective $U$ value the energy dispersions,
Eqs. (\ref{epsilon-cq}) and (\ref{epsilon-sq}), are not affected by the renormalisation that accounts for
the short-range interactions. As reported above, the effective value $U=0.8\,t$ is determined
by the ratio $W_{\rm h}/W_s$ of the experimentally observed ${\rm c}$ band (holon) and ${\rm s}$ band (spinon) energy bandwidths 
$W_{\rm h}=\varepsilon_{\rm c} (2k_{\rm F}) - \varepsilon_{\rm c} (0)$ and $W_s =\varepsilon_{\rm s} (k_{\rm F}) - \varepsilon_{\rm c} (0)$,
respectively. Indeed, within the 1D Hubbard model the $W_{\rm h}/W_s$ value only depends on $U/t$ and
the electronic density $n$. For $n=2/3$ the agreement with the observed energy bandwidths is then
found to be reached for $U/t=0.8$.

However, the renormalisation fixes the effective $U$ value yet does not affect $t$.
This is due to symmetry implying that within the 1D Hubbard model the full ${\rm c}$ band
energy bandwidth $\varepsilon_{\rm c} (\pi) - \varepsilon_{\rm c} (0)$ is independent of $U$ and $n$ and
exactly reads $4t$. That energy bandwidth can be written as $W_{\rm h} + W_{\rm c} = 4t$ where for the
present metallic phase the energy bandwidth $W_{\rm c} = \varepsilon_{\rm c} (\pi) - \varepsilon_{\rm c} (2k_{\rm F})$ is finite. 
Within our pseudofermion representation, $W_{\rm h}$ and $W_{\rm c}$ are the ${\rm c}$ band filled 
and unfilled, respectively, ground-state Fermi sea energy bandwidths. Again, the value of the ratio $W_{\rm h}/W_{\rm c}$ 
only depends on $U/t$ and the electronic density $n$. Accounting for the $W_{\rm h}/W_{\rm c}$ value at $U/t=0.8$ and 
$n=2/3$ together with the exact relation $W_{\rm h} + W_{\rm c} = 4t$ one finds from analysis of Figs. 5 (d)-(f) that 
$t \approx 0.58$\,eV for the MoSe$_2$ 1D quantum-line defects.

Such defects experimental uncertainty interval $\alpha\in[0.75,0.80]$ of the exponent that controls 
the low-$\omega$ electronic density of states suppression $\vert\omega\vert^\alpha$ is outside the corresponding 
1D Hubbard model range, Eq. (\ref{alphaHM}). Hence the $U = 0.8\,t$ value obtained from matching the corresponding 
ARPES cusps lines spectra with those of the 1D Hubbard model for electronic density $n=2/3$ refers to an 
effective interaction having contributions both from electron onsite and finite-range interactions. In addition to the 
interaction $U$ renormalisation, both the parameter $\xi_{\rm c}$ and the corresponding ${\rm c}$ pseudofermion phase shifts 
$2\pi\Phi_{c,\beta'}(\iota 2k_{\rm F},q_{{\rm F}\beta'})$ in Eq. (\ref{PhiFP}) where $\beta'= {\rm c,s}$ whose expressions involve $\xi_{\rm c}$
undergo a second renormalisation. It is such that $\xi_{\rm c}$ is replaced by a parameter ${\tilde{\xi}}_{\rm c}$ associated 
with $\alpha $ values in the range $\alpha \in [\alpha_0,\alpha_{\rm max}]$.

The universality referring to low-energy values in the vicinity of the ${\rm c}$ and ${\rm s}$ bands Fermi points 
implies that for the non-integrable model with finite-range interactions the relation $\alpha_0 = (2-\xi_{\rm c}^2)^2/8\xi_{\rm c}^2$ 
given in Eq. (\ref{alphaHM}) remains having the same form for $\alpha \in [\alpha_0,\alpha_{\rm max}]$ and 
${\tilde{\xi}}_{\rm c} \in [1/2,\xi_{\rm c}]$, so that,
\begin{equation}
\alpha = {(2-{\tilde{\xi}}_{\rm c}^2)^2\over 8{\tilde{\xi}}_{\rm c}^2} \, ; \hspace{0.35cm}
{\tilde{\xi}}_{\rm c} = \sqrt{2\Bigl(1 + 2\alpha - 2\sqrt{\alpha (1+\alpha)}\Bigr)} \, .
\label{alphaRHMgen}
\end{equation}
(The first equation other mathematical solution, ${\tilde{\xi}}_{\rm c} = \sqrt{2(1 + 2\alpha + 2\sqrt{\alpha (1+\alpha)})}$,
is not physically acceptable.) 

On the one hand, the spin $SU(2)$ symmetry imposes that the values of the $U/t$-independent parameter 
$\xi_s = \sqrt{2}$ and ${\rm s}$ pseudofermion phase shifts $2\pi\Phi_{s,\beta'}(\iota k_{\rm F},q_{{\rm F}\beta'})$ in Eq. (\ref{PhiFP}) where 
$\beta' = {\rm c,s}$ remain unchanged for the model with finite-range interactions. On the other hand, the general relations,
Eq. (\ref{PhiFP}), are universal so that for that model corresponding to any $\alpha $ value in the range 
$\alpha \in [\alpha_0,\alpha_{\rm max}]$ the ${\rm c}$ pseudofermion phase shifts 
$2\pi\Phi_{c,\beta'}(\iota 2k_{\rm F},q_{{\rm F}\beta'})$ are for $\beta'= {\rm c,s}$ given by,
\begin{eqnarray}
& & 2\pi{\tilde{\Phi}}_{\rm c,c}(\iota 2k_{\rm F},2k_{\rm F}) = \iota\,2\pi{\tilde{\Phi}}_{\rm c,c}(2k_{\rm F},\iota 2k_{\rm F}) 
\nonumber \\
& & = {\pi\,({\tilde{\xi}}_{\rm c} -1)^2\over {\tilde{\xi}}_{\rm c}}\hspace{0.35cm}{\rm for}\hspace{0.25cm}\iota = +1 \, ,
\nonumber \\
& & = - {\pi\,({\tilde{\xi}}_{\rm c}^2 -1)\over {\tilde{\xi}}_{\rm c}}\hspace{0.35cm}{\rm for}\hspace{0.25cm}\iota = -1 \, ,
\nonumber \\
& & 2\pi{\tilde{\Phi}}_{\rm c,s}(\iota 2k_{\rm F},k_{\rm F}) = \iota\,2\pi{\tilde{\Phi}}_{\rm c,s}(2k_{\rm F},\iota k_{\rm F}) 
\nonumber \\
& & = {\pi\over 2}\,{\tilde{\xi}}_{\rm c}
\hspace{0.35cm}{\rm for}\hspace{0.25cm}\iota = \pm 1 \, .
\label{RPhiFP}
\end{eqnarray}

The universality on which our scheme relies refers both to the low-energy TLL limit 
and to energy windows near the high-energy ${\rm c}$, ${\rm c'}$ and ${\rm s}$ branch-lines singularities. 
The expression of the exponents that control the spectral function behaviour at low energy and in
the vicinity of such singularities only involves the phase shifts of ${\rm c}$ and ${\rm s}$ pseudofermions 
with momenta at their Fermi points $q = \pm 2k_{\rm F}$ and $q' = \pm k_{\rm F}$, respectively.
On the one hand, as result in part of the spin $SU(2)$ symmetry, at zero spin density the general ${\rm s}$ 
pseudofermion phase shifts $2\pi{\tilde{\Phi}}_{\rm s,s}(q',q)$ and $2\pi{\tilde{\Phi}}_{\rm s,c}(q',q)$ 
remain unchanged for their whole momentum intervals. On the other hand, the general phase shifts 
$2\pi{\tilde{\Phi}}_{\rm c,c}(q,q')$ and $2\pi{\tilde{\Phi}}_{\rm c,s}(q,q')$ of ${\rm c}$ pseudofermions whose momenta 
have absolute values $\vert q\vert<2k_{\rm F}$ inside the ${\rm c}$ band Fermi sea contribute neither to the TLL 
low-energy spectral function expression nor to the high-energy branch-lines exponents. 
Consistently, similarly to the ${\rm s}$ pseudofermion phase shifts  $2\pi{\tilde{\Phi}}_{\rm s,s}(q',q)$ and $2\pi{\tilde{\Phi}}_{\rm s,c}(q',q)$,
they remain unchanged upon increasing $\alpha$ from $\alpha = \alpha_0$.

Hence the main issue here is the renormalisation of phase shifts of ${\rm c}$ pseudofermions with momenta at 
the Fermi points, $2\pi{\tilde{\Phi}}_{\rm c,c}(\iota 2k_{\rm F},q)$ and $2\pi{\tilde{\Phi}}_{\rm c,s}(\iota 2k_{\rm F},q')$
for $\iota =\pm 1$. Multiplying $2\pi{\tilde{\Phi}}_{\rm c,c}(\iota 2k_{\rm F},q)$ and $2\pi{\tilde{\Phi}}_{\rm c,s}(\iota 2k_{\rm F},q')$ 
by the phase factor $-1$ gives the phase shifts acquired by the ${\rm c}$ pseudofermions of
momenta $q=\iota 2k_{\rm F} = \pm 2k_{\rm F}$ upon scattering off one ${\rm c}$ band hole (holon) created under a transition 
to an excited state at any momentum $q$ in the interval $q \in [-2k_{\rm F},2k_{\rm F}]$ 
and one ${\rm s}$ band hole (spinon) created at any momentum $q'$ in the domain $q' \in [-k_{\rm F},k_{\rm F}]$, respectively. 
The overall phase-shift renormalisation must preserve the ${\rm c}$ pseudofermion phase-shifts values given in Eq. (\ref{RPhiFP}) 
for (i) $q=\iota 2k_{\rm F}=\pm 2k_{\rm F}$ and (ii) $q'=\iota k_{\rm F}=\pm k_{\rm F}$. Hence it introduces suitable factors 
multiplying $2\pi\Phi_{\rm c,c}(\iota 2k_{\rm F},q)$ and $2\pi\Phi_{\rm c,s}(\iota 2k_{\rm F},q')$.
In the case of $2\pi{\tilde{\Phi}}_{\rm c,c}(\iota 2k_{\rm F},q)$, this brings about a singular behaviour at $q = -\iota 2k_{\rm F}$
for $\alpha>\alpha_0$ similar to that in the ${\rm s}$ pseudofermion phase shift 
$2\pi\Phi_{\rm s,s}(\iota k_{\rm F},q')$ at $q' = \iota k_{\rm F}$, supplementary Equation (15),
for the conventional 1D Hubbard model, which remains having the
same values for the renormalised model.

The ${\rm c}$ and ${\rm s}$ pseudofermion phase shifts of the 1D Hubbard model with electron finite-range interactions
are for the whole range $\alpha \in [\alpha_0,\alpha_{\rm max}]$ thus of the general form,
\begin{eqnarray}
2\pi{\tilde{\Phi}}_{\rm c,c}(q,q') & = & 2\pi\Phi_{\rm c,c}(q,q')  
\hspace{0.15cm}{\rm for}\hspace{0.1cm}q\neq \iota 2k_{\rm F} \, , \hspace{0.1cm} \iota = \pm 1 \, ,
\nonumber \\
2\pi{\tilde{\Phi}}_{\rm c,c}(\iota 2k_{\rm F},q') & = & {\xi_{\rm c}\over {\tilde{\xi}}_{\rm c}}{({\tilde{\xi}}_{\rm c} -1)({\tilde{\xi}}_{\rm c} -(-1)^{\delta_{q',-\iota 2k_{\rm F} }})
\over (\xi_{\rm c} -1)(\xi_{\rm c} -(-1)^{\delta_{q',-\iota 2k_{\rm F} }})}
\nonumber \\
& \times & 2\pi\Phi_{\rm c,c}(\iota 2k_{\rm F},q)  
\hspace{0.15cm}{\rm for}\hspace{0.1cm}\iota = \pm 1 \, ,
\nonumber \\
2\pi{\tilde{\Phi}}_{\rm c,s}(q,q') & = & 2\pi\Phi_{\rm c,s}(q,q') \hspace{0.15cm}{\rm for}\hspace{0.1cm}q\neq \iota 2k_{\rm F} \, , \hspace{0.1cm} \iota = \pm 1 \, ,
\nonumber \\
2\pi{\tilde{\Phi}}_{\rm c,s}(\iota 2k_{\rm F},q') & = & {{\tilde{\xi}}_{\rm c}\over\xi_{\rm c}}\,2\pi\Phi_{\rm c,s}(\iota 2k_{\rm F},q') \hspace{0.15cm}{\rm for}\hspace{0.1cm}\iota = \pm 1 \, ,
\nonumber \\
2\pi{\tilde{\Phi}}_{\rm s,s}(q',q) & = & 2\pi\Phi_{\rm s,s}(q',q) \, ,
\nonumber \\
2\pi{\tilde{\Phi}}_{\rm s,c}(q',q) & = & 2\pi\Phi_{\rm s,c}(q',q) \, . 
\label{Phi-rela-MoSe2}
\end{eqnarray}

Our theoretical results refer to the thermodynamic limit at $T=0$. In that case the phase-shifts renormalisation,
Eq. (\ref{Phi-rela-MoSe2}), only affects those of the ${\rm c}$ pseudofermion scatterers with momentum values $\pm 2k_{\rm F}$ corresponding
to the zero-energy Fermi level. Note however that the corresponding ${\rm c}$ and ${\rm s}$ pseudofermion scattering centers
have momenta $q\in [-2k_{\rm F},2k_{\rm F}]$ and $q'\in [-k_{\rm F},k_{\rm F}]$, respectively, 
that correspond to a large range of high-energy values. At finite temperature 
$T\approx 300$\,K one has that $k_B\,T\approx 0.045\,t$ where $t \approx 0.58$\,eV is within the present 
theoretical description the transfer integral value suitable for the MoSe$_2$ 1D quantum-line defects.
The derivation of some of the theoretical expressions involves a $T=0$ ${\rm c}$ band momentum distribution that 
reads one for $\vert q\vert <2k_{\rm F}$ and zero for $2k_{\rm F}<\vert q\vert <\pi$. At finite temperature 
$T\approx 300$\,K, such a distribution is replaced by a ${\rm c}$ pseudofermion
Fermi-Dirac distribution. This implies for instance that the $q=\pm 2k_{\rm F}$ ${\rm c}$ pseudofermion 
phase-shift renormalisation in Eq. (\ref{Phi-rela-MoSe2}) is extended from the zero-energy Fermi level 
to a small region of energy bandwidth $0.045\,t\approx 0.026$\,eV near the ${\rm c}$ band Fermi points 
$q=\pm 2k_{\rm F}$. This refers to a corresponding small region with the same energy bandwidth 
near the physical Fermi points $k=\pm k_{\rm F}$ in Fig. \ref{figure5} (d), (e) and (f).
Interestingly, finite-size effects have at $T=0$ the similar effect of slightly enhancing 
the energy bandwidth of the ${\rm c}$ pseudofermion phase shifts renormalisation, Eq. (\ref{Phi-rela-MoSe2}), 
in the very vicinity of the zero-energy Fermi level. Hence any small finite temperature and/or the system
finite size remove/s the singular behaviour of the phase-shifts renormalisation being restricted to the 
zero-energy Fermi level.

Fortunately, both the finite size of the MoSe$_2$ 1D quantum-line defects and
the experimental temperature $\approx 300$\,K  lead though to very small effects, as confirmed by 
the quantitative agreement reached between the $T=0$ theoretical results associated with the 
1D Hubbard model with electron finite-range interactions and the  experimental data. Hence
for simplicity in the following we remain using our $T=0$ theoretical analysis in
terms of that model in the thermodynamic limit.\\ \\
{\bf Spectral function accounting for finite-range interactions.}
For energy windows corresponding to small $\gamma = {\rm c,c',s}$ energy deviations 
$(\omega_{\gamma} (k)-\omega)>0$ from the high-energy branch-line spectra 
$\omega_{\gamma} (k)$ given in Eqs. (\ref{omegac})-(\ref{omegas}), which as confirmed below
remain unchanged upon increasing $\alpha$ from $\alpha_0$, the general form of the electron removal
spectral function, Eq. (\ref{BHM}), and corresponding exponent, Eq. (\ref{3expoH}),
prevails for the model with finite-range interactions corresponding to $\alpha \in [\alpha_0,\alpha_{\rm max}]$. 
Hence for these energy windows that spectral function has the same universal form as in Eq. (\ref{BHM}),
\begin{equation}
B (k,\omega) \propto (\omega_{\gamma} (k)-\omega)^{{\tilde{\zeta}}_{\gamma} (k)} 
\hspace{0.35cm}{\rm for}\hspace{0.25cm}\gamma = {\rm c,c',s} \, .
\label{BHMFA}
\end{equation}
Both within the PDT ($\alpha = \alpha_0$) and
RPDT ($\alpha > \alpha_0$), most of the one-electron spectral weight is located in
the $(k,\omega)$-plane at and near the singular branch lines. Those refer to the $k$ ranges
of the $\gamma = {\rm c,c',s}$ branch lines for which the corresponding exponent ${\tilde{\zeta}}_{\gamma} (k)$
in Eq. (\ref{BHMFA}) is negative. For further information on the validity of the spectral functions expressions, 
Eqs. (\ref{BHM}) and (\ref{BHMFA}), and the definition of some quantities used in our theoretical analysis, see
supplementary Note 3.

We start by confirming that the ${\rm c}$ and ${\rm s}$ pseudofermion energy dispersions in the expressions of the 
$\gamma = {\rm c,c',s}$ branch-lines spectra $\omega_{\gamma} (k)$, Eqs. (\ref{omegac})-(\ref{omegas}), remain unchanged. 
This follows from the behaviour of the phase shifts appearing in these pseudofermion energy
dispersions expressions, Eqs. (\ref{epsilon-cq}) and (\ref{epsilon-sq}). In the
case of the conventional 1D Hubbard model, the integral $\int_{-Q}^{Q}dk$ over 
the rapidity momentum $k$ in the integrand rapidity phase shifts $2\pi\bar{\Phi }_{\rm c,c}
\left(\sin k/u, \sin k (q)/u\right)$ and $2\pi\bar{\Phi }_{\rm c,s}\left(\sin k/u, \Lambda (q')/u\right)$ of
Eqs. (\ref{epsilon-cq}) and (\ref{epsilon-sq}) can be transformed into a momentum integral
$\int_{-2k_{\rm F}}^{2k_{\rm F}}dq''$ over the whole ${\rm c}$ band Fermi sea with the integration 
momentum $q''\in [-2k_{\rm F},2k_{\rm F}]$ appearing in corresponding
integrand ${\rm c}$ pseudofermion phase shifts $2\pi\Phi_{\rm c,c} (q'',q)$ and $2\pi\Phi_{\rm c,s} (q'',q')$, respectively.

Under the electron finite-range interactions renormalisation, the latter phase shifts become 
$2\pi{\tilde{\Phi}}_{\rm c,c} (q'',q)$ and $2\pi{\tilde{\Phi}}_{\rm c,s} (q'',q')$, respectively, as defined in Eq. (\ref{Phi-rela-MoSe2}).
As given in that equation, the latter ${\rm c}$ pseudofermion phase shifts are only renormalised at the Fermi points, 
$q'' = \pm 2k_{\rm F}$. Hence such phase shifts renormalised values refer only
to the limiting values of the integration $\int_{-2k_{\rm F}}^{2k_{\rm F}}dq''$. 
The phase-shift contributions associated with such limiting momentum values $-2k_{\rm F}$ and
$+2k_{\rm F}$ have in the thermodynamic limit vanishing measure relative to the phase-shift
contributions from the range $-2k_{\rm F}<q''<2k_{\rm F}$ in $\int_{-2k_{\rm F}}^{2k_{\rm F}}dq''$. 
For $\vert q''\vert<2k_{\rm F}$ the phase shifts $2\pi{\tilde{\Phi}}_{\rm c,c} (q'',q)$ and $2\pi{\tilde{\Phi}}_{\rm c,s} (q'',q')$ remain unchanged,
see Eq. (\ref{Phi-rela-MoSe2}). Hence the energy dispersions 
$\varepsilon_{\rm c} (q) = \varepsilon_{\rm c}^0 (q) - \varepsilon_{\rm c}^0 (2k_{\rm F})$, Eq. (\ref{epsilon-cq}),
and $\varepsilon_{\rm s} (q) = \varepsilon_{\rm s}^0 (q) - \varepsilon_{\rm s}^0 (k_{\rm F})$, Eq. (\ref{epsilon-sq}),
remain as well unchanged. The same thus applies to the $\gamma = {\rm c,c',s}$ spectra $\omega_{\gamma} (k)$,
Eqs. (\ref{omegac})-(\ref{omegas}), in the spectral function expression, Eq. (\ref{BHMFA}).

In contrast, one finds from the combined use of Eqs. (\ref{3expoH}) and (\ref{Phi-rela-MoSe2}) that for the model with finite-range interactions the 
momentum dependent exponents in that expression are renormalised. For $U/t>0$, electronic densities $n\in [0,1]$ 
and $\alpha \in [\alpha_0,\alpha_{\rm max}]$ they are given by,
\begin{eqnarray}
{\tilde{\zeta}}_{\rm c} (k) & = & -{1\over 2} + \sum_{\iota=\pm1}\left({{\tilde{\xi}}_{\rm c}\over 4} + {\rm sgn}\{k\}{\tilde{\Phi}}_{\rm c,c}(\iota 2k_{\rm F},q)\right)^2  
\nonumber \\
k & = & \in [-k_{\rm F},k_{\rm F}] \, , 
\nonumber \\
q & = & -{\rm sgn}\{k\} k_{\rm F} - k \in [-2k_{\rm F},-k_{\rm F}]\,;[k_{\rm F},2k_{\rm F}] \, ,
\nonumber \\
{\tilde{\zeta}}_{\rm c'} (k) & = & -{1\over 2} + \sum_{\iota=\pm1}\left({{\tilde{\xi}}_{\rm c}\over 4} - {\rm sgn}\{k\}{\tilde{\Phi}}_{\rm c,c}(\iota 2k_{\rm F},q)\right)^2  
\nonumber \\
k & = & \in [-3k_{\rm F},3k_{\rm F}] \, , 
\nonumber \\
q & = & {\rm sgn}\{k\} k_{\rm F} - k \in [-2k_{\rm F},k_{\rm F}]\,;[-k_{\rm F},2k_{\rm F}] \, .
\nonumber \\
{\tilde{\zeta}}_{\rm s} (k) & = & -1 + \sum_{\iota=\pm1}\left({\iota\over 2{\tilde{\xi}}_{\rm c}} + {\tilde{\Phi}}_{\rm c,s} (\iota 2k_{\rm F},q')\right)^2  
\nonumber \\
k & \in & [-k_{\rm F},k_{\rm F}]\hspace{0.25cm}{\rm and}\hspace{0.25cm}q' = -k \in  [-k_{\rm F},k_{\rm F}] \, .
\label{3expoRH}
\end{eqnarray}

Plotting the momentum dependence of these exponents requires again the use of
exact numerical methods to solve the corresponding sets of coupled integral
equations. The momentum dependences found from that exact numerical solution are
plotted in Fig. \ref{figure7} as a function of the momentum $k$ for $U/t=0.8$, $t=0.58$\,eV, $n=2/3$ 
and representative $\alpha$ values $\alpha =\alpha_0\approx 1.4\times 10^{-3}$,
$\alpha = 0.70$, $\alpha = 0.7835 \approx 0.78$ and $\alpha = 0.85$. Their choice is confirmed 
below to be suitable for the discussion of the relation between the theoretical results and the observed 
spectral features.

The physics associated with the $\alpha$ range $\alpha \in [\alpha_0,1/8]$ is qualitatively different from
that corresponding to $\alpha \in [1/8,\alpha_{\rm max}]$. Note that at $\alpha =1/8$ and thus ${\tilde{\xi}}_{\rm c}=1$
the ${\rm c}$ pseudofermion phase shift $2\pi{\tilde{\Phi}}_{\rm c,c}(\iota 2k_{\rm F},q)$ in Eq. (\ref{Phi-rela-MoSe2}) exactly vanishes.
This vanishing marks the transition between the two physical regimes. The ${\rm c}$ pseudofermion phase shift 
$2\pi\Phi_{\rm c,c}(\iota 2k_{\rm F},q)$ of the conventional 1D Hubbard model also vanishes in the limit of infinity onsite repulsion
in which $\alpha_0=1/8$. Increasing $\alpha$ from $\alpha = \alpha_0$ within the interval $\alpha \in [\alpha_0,1/8]$ indeed
increases the actual onsite repulsion, which for $\alpha>\alpha_0$ is not associated 
anymore with the renormalised model constant effective $U$ value. In addition, it introduces electron finite-range interactions. 
On the one hand, in that $\alpha$ interval the effects on the $\gamma = {\rm c,c',s}$ exponents, Eq. (\ref{3expoRH}), of increasing 
$\alpha$ are controlled by the increase of the actual onsite repulsion. On the other hand, as $\alpha$ changes within the
interval $\alpha \in [\alpha_0,1/8]$ the fixed effective $U$ value accounts for both effects from the
actual onsite interaction and emerging finite-range interactions. It imposes that 
the ${\rm c}$ and ${\rm s}$ pseudofermion energy dispersions in Eqs. (\ref{epsilon-cq}) and (\ref{epsilon-sq})
remain as for that $U$ value. This means
that the effects of increasing the actual onsite repulsion due to increasing $\alpha$ are on the matrix
elements of the electron annihilation operator between energy eigenstates that control the branch-lines 
exponents, Eq. (\ref{3expoRH}), and thus the spectral weights.

For $U/t=0.8$, $t=0.58$\,eV and $n=2/3$ the ${\rm c}$, ${\rm c'}$ and ${\rm s}$ branch-lines exponents, Eq. (\ref{3expoRH}), 
corresponding to $\alpha=1/8$ are represented in Fig. \ref{figure7} (a), (b) and (c), respectively, by the dotted lines. 
The changes in these exponents caused by increasing the $\alpha$ 
value from $\alpha_0$ to $1/8$ relative to the exponents curves given for the $\alpha_0\approx 1.4\times 10^{-3}$ conventional 1D Hubbard model 
in that figure are qualitatively similar to those originated by increasing $U/t$ from $0.8$ to infinity
within the latter model. Such an increase also enhances $\alpha_0$ from 
$\alpha_0\approx 1.4\times 10^{-3}$ to $1/8$. The main difference relative
to the conventional 1D Hubbard model is that the ${\rm c}$ and ${\rm s}$ pseudofermion energy dispersions remain unchanged upon
increasing $\alpha$. Comparison of the momentum intervals of the $\gamma = {\rm c,c',s}$ branch lines
for which the exponents, Eq. (\ref{3expoRH}), are negative for $\alpha \in [\alpha_0,1/8]$ 
with those in which there are cusps in the experimental dispersions of Figs. 5 (e) and (f) reveals that there is no agreement 
between theory and experiments for that $\alpha$ range.

Further increasing $\alpha$ within the interval $\alpha \in ]1/8,\alpha_{\rm max}]$ corresponds
to a different physics. The changes in the branch-lines exponents, Eq. (\ref{3expoRH}), are then mainly 
due to the increasing effect of the electron finite-range interactions
upon increasing $\alpha$. It leads in general to a corresponding increase of 
the three $\gamma = {\rm c,c',s}$ exponents ${\tilde{\zeta}}_{\gamma} (k)$, Eq. (\ref{3expoRH}).
For $U/t=0.8$, $t=0.58$\,eV, $n=2/3$ and both $\alpha \in ]1/8,0.75]$ and $\alpha \in [0.78,\alpha_{\rm max}]$
the momentum intervals of the $\gamma = {\rm c,c',s}$ branch lines
for which these exponents are negative do not agree to those for which there are cusps in the
MoSe$_2$ 1D quantum-line defects measured spectral function. In order to illustrate the $\alpha$ dependence
of the $\gamma = {\rm c,c',s}$ branch lines exponents, Eq. (\ref{3expoRH}), their $k$ dependence has been plotted
in Fig. \ref{figure7} for the set of representative $\alpha$ values $\alpha =\alpha_0\approx 1.4\times 10^{-3}$,
$\alpha = 0.70$, $\alpha = 0.7835\approx 0.78$ and $\alpha = 0.85$.

The following analysis refers again to the values $U/t=0.8$, $t=0.58$\,eV and $n=2/3$ associated with
the MoSe$_2$ 1D quantum-line defects. For $\alpha < 0.75$ the momentum
width of the $\gamma =c'$ branch line $k$ range for which its exponent ${\tilde{\zeta}}_{\rm c'} (k)$ is negative
is larger than that of the experimental dispersion shown in Figs. 5 (e) e (f) near the corresponding
excitation energy $\approx 0.95$\,meV. Upon increasing $\alpha$ from $\alpha=0.75$, the
$\gamma =c'$ branch line momentum width for which ${\tilde{\zeta}}_{\rm c'} (k)$ is negative
continuously decreases, vanishing at $\alpha = 0.7835\approx 0.78$.
Comparison of the momentum ranges for which the exponents plotted in Fig. \ref{figure7} are negative with those
in which there are cusps in the experimental dispersions of Figs. \ref{figure5} (e) e (f) reveals
that there is quantitative agreement for $\alpha \in [0.75,0.78]$. Further increasing $\alpha$ from
$\alpha =0.78$ leads to a ${\rm c}$ branch line momentum width around $k=0$ in which the
exponent ${\tilde{\zeta}}_{\rm c} (k)$ becomes positive. This disagrees with the observation of
experimental cusps near the excitation energy $\approx 0.85$\,meV around $k=0$ and for
decreasing energy along the ${\rm c}$ branch line upon further increasing $\alpha$.

That there is quantitative agreement between theory and
the experiments for $\alpha \in [0.75,0.78]$ is fully consistent with the corresponding
$\alpha$ uncertainty range $\alpha \in [0.75,0.80]$ found independently 
from the DOS suppression experiments.
The momentum dependence of the $\gamma = {\rm c,c',s}$ branch lines exponents corresponding
to $\alpha =0.78$ is represented by full lines in Figs. 5 (a), (c) and (d) for $U=0.8\,t$,
$t=0.58$\,eV and electronic density $n=2/3$. 

As for the exponents expressions, Eq. (\ref{3expoH}), those of the ${\rm c}$ and ${\rm s}$ branch-line exponents given in
Eq. (\ref{3expoRH}) are not valid at the low-energy limiting values $k=\pm k_{\rm F}$. While in the
thermodynamic limit this refers only to $k=\pm k_{\rm F}$, for the finite-size MoSe$_2$ 1D quantum-line defects
it may refer to two small low-energy regions in the vicinity of $k=\pm k_{\rm F}$. Both this property and the
positivity of the ${\rm s}$ branch exponent for $\alpha \in [0.75,0.78]$ in these momentum
regions are consistent with the lack of low-energy cusps in the ARPES data shown in Fig. \ref{figure5} (e) and (f).

We have calculated the $k$ and $\omega$ dependence of the spectral function expression of the 1D Hubbard model with finite-range
interactions near the ${\rm c}$ and ${\rm s}$ branch lines in the momentum ranges for which they display cusps, Eq. (\ref{BHMFA}). 
If one goes away from the $(k,\omega)$-plane vicinity of these lines, one confirms that both such a model spectral 
function and that of the conventional 1D Hubbard model have the broadening discussed in the supplementary Note 2.

For a short discussion on whether the RPDT is useful to extract information beyond that given by the conventional 
1D Hubbard model and corresponding PDT about the physics of quasi-1D metals and a comparison of the PDT and 
RPDT theoretical descriptions of the line defects in MoSe$_2$, see the supplementary Note 4.\\ \\ 
{\bf Data Availability}\\ 
The datasets generated during and/or analyzed during the current study are available from the corresponding authors on reasonable request.

%%%%%%%%%%%%%%%%%%%%%%%%%%%%%%%%%%%%%%%%%%%%%%%%%%%%%%%%%%%%%%%%%%%%%%%%%%
 
%%%%%%%%%%%%%%%%%%%%%%%%%%%%%%%%%%%%%%%%%%%%%%%%%%%%%%%%%%%%%%%%%%%%%%%%%%

{\bf Acknowledgments}\\
The USF group acknowledges support from the National Science Foundation (DMR-1204924). V.K., R.D. and M.-H. P. 
acknowledges support from the Army Research Office (W911NF-15-1-0626) and thank Prof. Hari Srikanth for 
resistance measurements in his laboratory. M.C.A., J.A. and C. C. thank enlightening exchanges with Grabriel Kotliar and Zhi-Xun Shen.
The Synchrotron SOLEIL is supported by the Centre National de la Recherche Scientifique (CNRS) and 
the Commissariat \`a l'Energie Atomique et aux Energies Alternatives (CEA), France. T.\v{C}. and J.M.P.C.
thank Eduardo Castro, Hai-Qing Lin and Pedro D. Sacramento for illuminating discussions. The theory group acknowledges the 
support from NSAF U1530401 and computational resources from CSRC (Beijing), the Portuguese FCT through the Grant 
UID/FIS/04650/2013, and the NSFC Grant 11650110443.\\ \\

{\bf Author Contributions}\\
Y.M. and H.C.D. contributed equally to this work. They both grew samples by MBE and characterized them by STM. The ARPES data have been obtained and analyzed by J.A., H.C.D., C.C. and M.C.A.. The 4-point transport measurements have been conducted and discussed by R.D., V.K. and M.-H.P. .
The project has been conceived by M.B. and M.C.A. who directed its experimental part. The theoretical description has been conceived by J.M.P.C. and the corresponding theoretical analysis was carried out
by T.\v{C}. and J.M.P.C.. The manuscript has been written by M.B., M.C.A. and J.M.P.C.. All authors contributed to the scientific discussion, contributed to and agreed on the manuscript.\\ \\
{\bf Additional information}\\ 
Supplementary information is available in the online version of the paper. 
Correspondence and requests for materials should be addressed to 
M.C.A. (ARPES experiments), M.B. (STM, growth, 4-point probe) and J.M.P.C. (theory).\\ \\
{\bf Competing Financial Interests}\\
The authors declare no competing financial interests.\\ \\

\clearpage

\onecolumngrid

\begin{figure}
\centerline{\includegraphics[width=14cm]{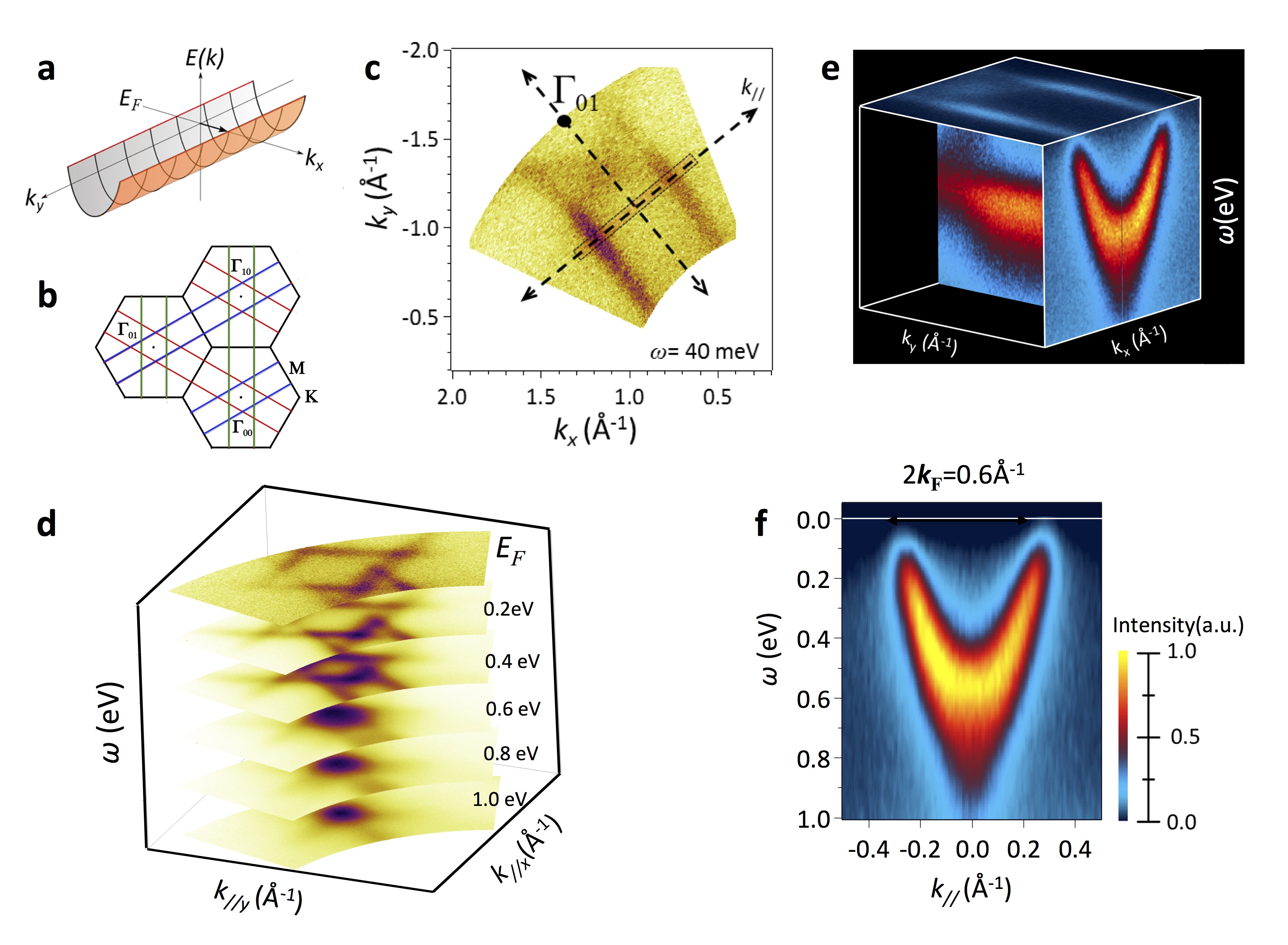}}
\caption{{\bf ARPES measurement of $k$-space resolved electronic structure of MTBs.} 
In (a) the band structure of a 1D metal 
is schematically illustrated. The parabolic band disperses in the $k_{\rm x}$ direction, which is the momentum vector along the 
1D defect. The lack of periodicity in the $k_{\rm y}$ direction causes the replication of the parabola forming a parabolic through 
and thus the Fermi-surface consists of two parallel lines. In the case of the 3 equivalent directions of MTBs that are rotated 
by 120$^{\circ}$ with respect to each other, three Fermi-surfaces overlap to form the Fermi-surface illustrated in (b). 
The experimental measurement of the Fermi-surface close to the center of the 2nd BZ using left and right circular polarized 
light is shown in (d). By using linear polarized light photoemission from a specific MTB-orientation can be emphasized as 
shown in (c). The Band dispersion $E (k)$ is shown in (e) and (f) for the momentum slice indicated in (c). This momentum 
slice was chosen because it lies outside of bands for the other two MTB orientations.}
\label{figure3}
\end{figure}

\begin{figure}
\centerline{\includegraphics[width=12cm]{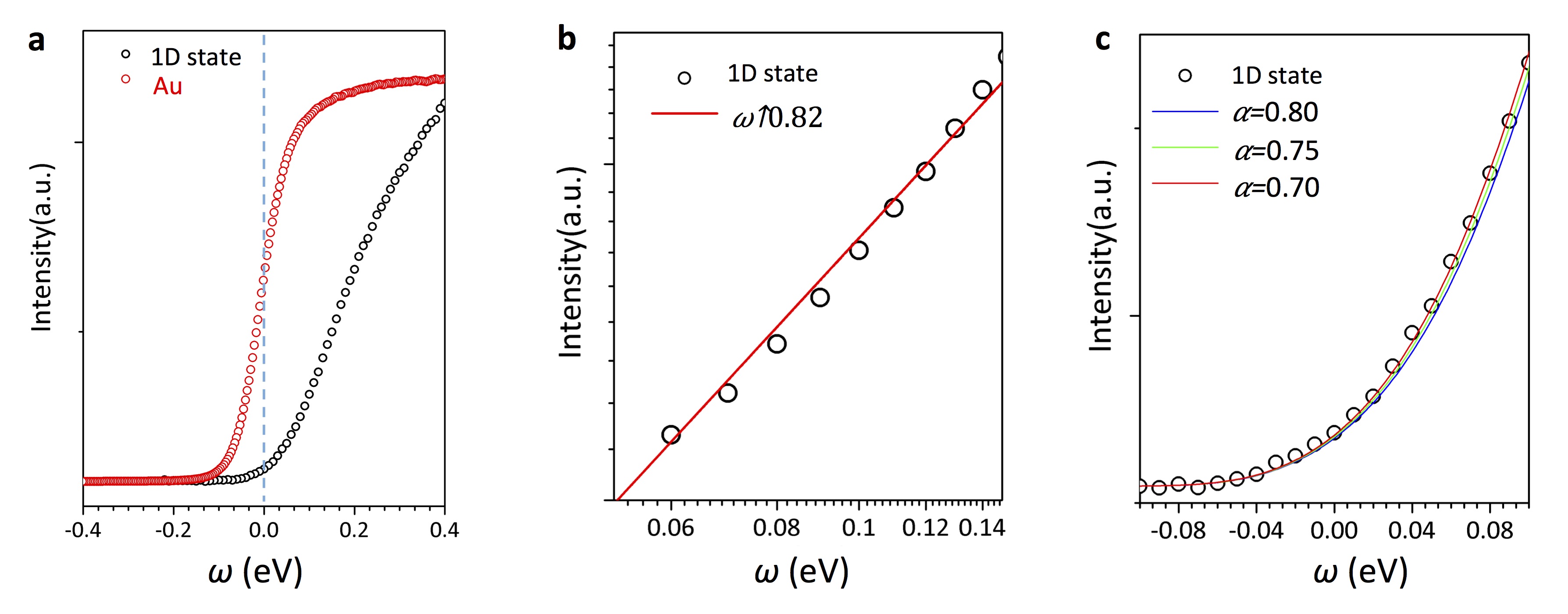}}
\caption{{\bf Evaluation of the suppression of the density of states at the Fermi level according to TLL theory.}
The suppression of the density of states of MTBs close to the Fermi-level compared to the density of states for a regular FL metal (Au) 
is shown in (a), measured at room temperature (to avoid CDW transition). The density of states is obtained by plotting the 
angle integrated photoemission intensity as a function of binding energy $\omega$. The log plot in (b) indicates that the density of 
states increases\cite{Schonhammer-93} with $\omega^{0.8}$, as is shown in (c). 
The data are well fit with $\alpha = 0.75$, but the variation of the fit with the exponent is small and thus the 
uncertainty in $\alpha$ is estimated to be as large as $\pm 0.05$.}
\label{figure4}
\end{figure}

\begin{figure}
\centerline{\includegraphics[width=11cm]{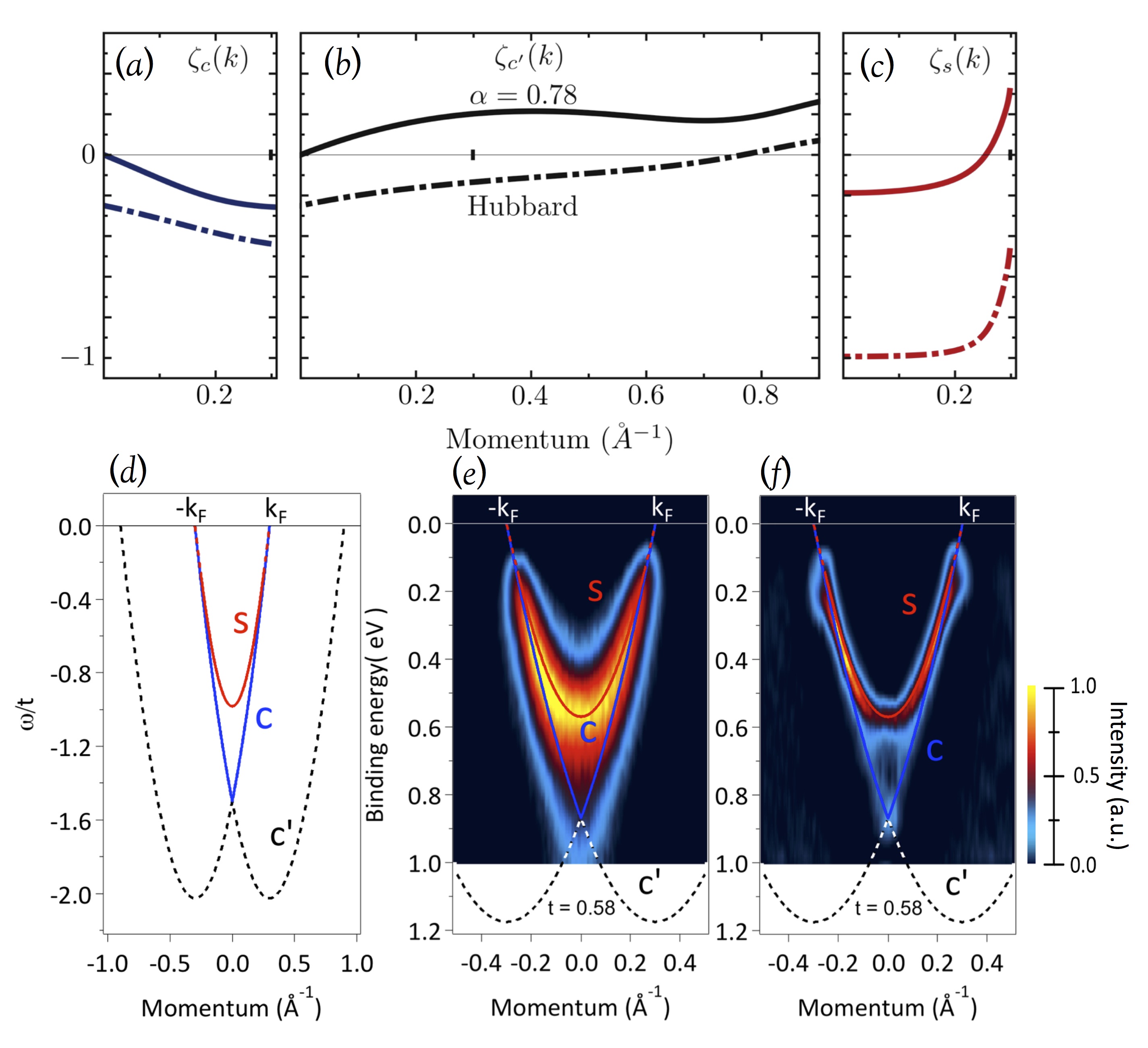}}
\caption{{\bf Exponents momentum dependence and theoretical and experimental
spectral lines.}
(a), (b) and (c) The exponents that control the spectral function near the ${\rm c}$, ${\rm c'}$ and ${\rm s}$ branch lines, respectively, 
for $U/t=0.8$, $t=0.58$\,eV and electronic density $n=2/3$ plotted as a function of $k$ for the 1D Hubbard model with finite-range 
interactions corresponding to $\alpha = 0.78$ (full lines) and the conventional 1D Hubbard model for which
$\alpha_0 \approx 1.4\times 10^{-3}$ (dashed-dotted lines), respectively. For the former model
at $\alpha = 0.78$ the ${\rm c'}$ branch line exponent remains positive for all its $k$ range whereas the ranges for which 
the ${\rm c}$ and ${\rm s}$ branch lines exponents are negative coincide with the momentum intervals showing 
ARPES peaks in (e) and (f); (d) The theoretical ${\rm c}$, ${\rm c'}$ and ${\rm s}$ branch line spectra plotted as a function of the 
momentum $k$ for the 1D Hubbard model with finite-range interactions corresponding to $\alpha = 0.78$ whose 
full and dashed lines refer to momentum ranges with negative and positive exponents, respectively; (e) Energy 
$vs$ momentum (k$_{//}$) along the $\overline{\Gamma_{01}}\,\overline{K}$ direction in the Brillouin zone, plus 
the same theoretical lines as in (d). The broad spectral line and the spectral continuum between the ${\rm s}$ and ${\rm c}$ branch lines 
apparent in (e) are consistent with the behaviour of 1D metals and our theoretical model, see Methods section and supplementary Note 2 for details. 
The results of applying a curvature procedure to the raw data$\cite{Zhang-11}$ 
on panel (e) are shown in panel (f), together with the theoretically computed ${\rm c}$, ${\rm c'}$ and ${\rm s}$ branch lines.}
\label{figure5}
\end{figure}

\begin{figure}
\centerline{\includegraphics[width=12cm]{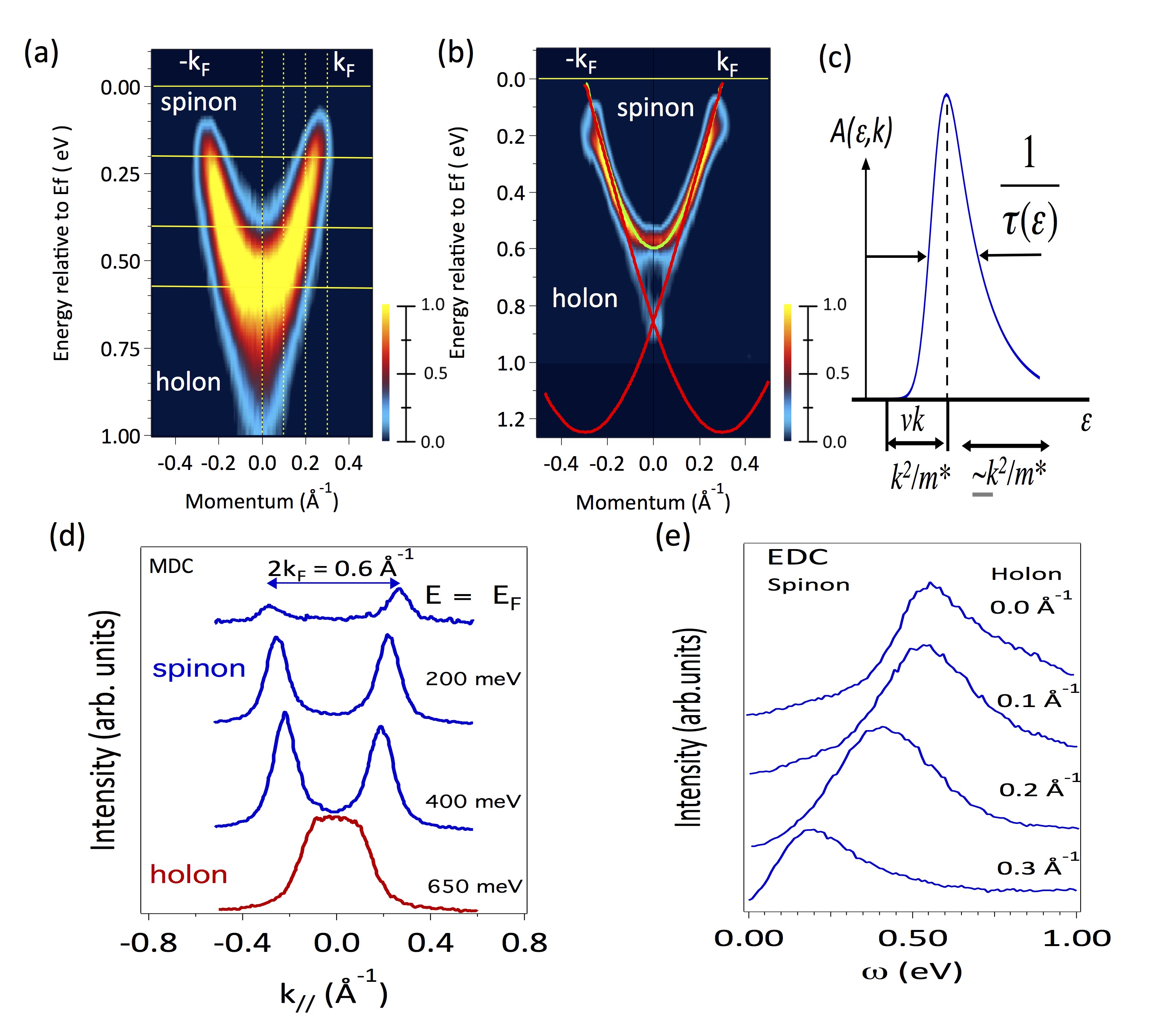}}
\caption{{\bf ARPES analysis using  EDC and MDC plots.} 
(a) Raw ARPES data, (b) second derivative of data in panel (a), (c) schematic description between the EDC shape and the lifetime, (d) MDC plots at different binding energies extracted from panel (a) data and finally panel (e) shows EDC plots at different momentum ($k$) values indicated in panel (a) as yellow straight lines.}
\label{figure6}
\end{figure}

\begin{figure}
\centerline{\includegraphics[width=12cm]{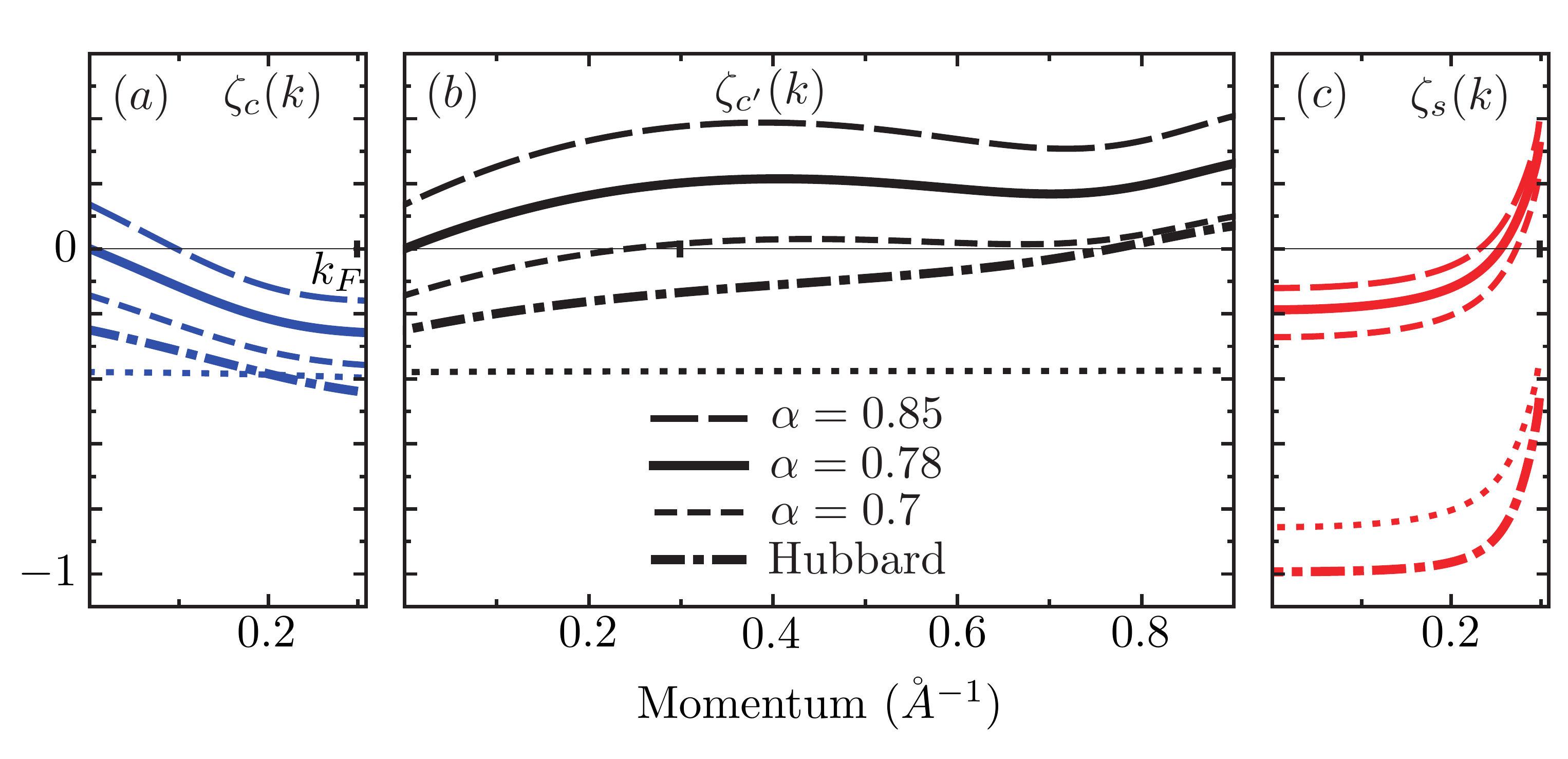}}
\caption{{\bf Momentum dependence of spectral-function exponents.}(a), (b) and (c): The ${\rm c}$, ${\rm c'}$ and ${\rm s}$ branch-lines exponents, respectively, defined in Eq. (\ref{3expoRH}) plotted as a function of the momentum $k$ for $U/t=0.8$, $t=0.58$\,eV, $n=2/3$ and representative $\alpha$ values 
$\alpha =\alpha_0\approx 1.4\times 10^{-3}$, $\alpha = 0.70$, $\alpha = 0.7835 \approx 0.78$
and $\alpha = 0.85$. In addition, the dotted lines refer to $\alpha =1/8$. As justified in the
text, for $\alpha\in [0.75,0.78]$ the momentum ranges of the ${\rm c}$, ${\rm c'}$ and ${\rm s}$ branch lines for which such exponents are 
negative coincide with those showing ARPES peaks in Fig. \ref{figure5} (e) and (f).}
\label{figure7}
\end{figure}		
		
\twocolumngrid
\end{document}

{\bf Supplementary Note 1: Characterisation of charge density wave  in mirror twin grain boundaries}\\ 

Charge density waves (CDW) are instabilities of the Fermi surface in the presence of electron-phonon coupling. The ideal case is when all electrons can be excited, which implies that all parts of Fermi surfaces are connected by the same $q$-vector. This property of a Fermi surface is called perfect nesting. For an ideal 1D system, the Fermi surface consists of two points (or two parallel lines), at $-k_{\rm F}$ and $+k_{\rm F}$ (where $k_{\rm F}$ is the Fermi wave vector), and thus exhibits perfect nesting at $q=2k_{\rm F}$. Hence, all one-electron (1D) systems are subject to CDW transitions for which this phenomenon is also known as Peierls transition. The periodicity of the charge density is given by $\pi/k_{\rm F}$. The Peierls transition is also often described by a change of the Brillouin zone (BZ) due to the change of the lattice periodicity to $\pi/k_{\rm F}$ i.e. the BZ boundary now coincides with $k_{\rm F}$. Therefore a band gap opens and the Peierls transition is associated with a metal-to-insulator transition. Fig. 2 (c) shows the characteristics of the CDW/Peierls transition.\\	

In the scanning tunnelling microscopy (STM) measurements the grain boundaries appear as two parallel lines\cite{Liu-14}, and can be assigned to the position of the Se$-$atoms neighbouring the defect line. The STM contrast is consistent with calculations\cite{Lehtinen-15} that show charge density expanding into vacuum at the Se atom positions adjacent to the defect. At room temperature, the measured corrugation along these Se-rows corresponds to the atomic lattice ($a_{{\rm MoSe}_2}$ = 3.3 \AA). Imaging the MTBs at lower temperatures causes their periodicity to increase to 3$\times a_{{\rm MoSe}_2}$, as shown in Figs. 2 (a) and (b). 
Such an increase in periodicity in STM is consistent with the formation of a charge density wave (CDW)\cite{Barja-16}. Thus the STM measurement suggests $k_{\rm F} = {\pi\over 3}{1\over a_{{\rm MoSe}_2}}\approx$ 0.32 \AA$^{-1}$ , which is in good agreement with Fermi-surface mapping by ARPES. The high resolution momentum resolved photoemission data show $k_{\rm F}/\pi$= 0.30\,\AA$^{-1}$.\\

The metal - insulator transition (MIT) associated with the CDW causes an increase of the electrical resistance of the sample. To probe this and to determine the transition temperature we performed simple four-point resistance measurements. For four point measurements a slightly thicker MoSe$_2$ film was grown of 1-2 monolayer thick to ensure connectivity between the MoSe$_2$ islands. Contacts to the sample were made by silver paint with contacts separated by $\approx$ 1mm. Standard van der Pauw geometry for four point measurements was adopted and the temperature dependent sample resistance was measured in a cryogenic probe station cooled with liquid nitrogen. During the measurement the sample current was kept constant and the potential drop was evaluated to determine the sample resistance. This procedure implies that as the sample resistance decreases at lower T the applied potential increases to maintain a constant current. For a critical applied potential a de-pinning of the CDW is then observed.

Consistent with MIT(s) our 4-point measurements, shown in Fig 2 (d), exhibit a resistance jump at $\approx$ 235 K and a second smaller increase at $\approx$ 200 K. The occurrence of multiple transitions is attributed to incommensurate and commensurate CDW transitions\cite{Sinchenko-13}. In addition, we also observed a drop of the resistance for a critical applied voltage (see Fig. 2 (d)). This behaviour is known from CDW in quasi-1D bulk crystals\cite{Gruner-88,Thorne-96}  and is caused by a de-pinning of the charge density wave from defects and a collective sliding of CDW. While CDW transition are expected for 
1D electron systems (1DES), it is no proof for it.\\ \\
{\bf Supplementary Note 2: Broadening of the ARPES spectral function and analysis of the EDC, MDC and lifetime}\\ 

It is well-know that in 1D metallic systems there are no quasi-particles in the vicinity of the Fermi surface. The excitations rather 
are gapless collective modes involving charge and spin degrees of freedom. All these features have dramatic consequences, 
particularly on the angle-resolved photoelectron spectroscopy (ARPES) directly measured spectral 
function of interacting 1D fermions. Specifically, any structure of the spectral 
function other than the Fermi-energy cut-off, is noticeably broad. This is another typical hallmark of spin-charge separation 
due to the fact that there are no stable excitations with the quantum numbers of the electron. The broadening of the peaks occurs 
both at low energy near the Fermi level and at higher energies. Effectively, this broadening 
of the photoemission peaks is a hallmark of the spin-charge separation.\\

All theses concepts have been robustly probed in the early ARPES time, see for more details Ref. [\cite{Valla-99}]. Due to the separation of charge and spin, one hole (or one electron) is always unstable to decay into two or more elementary excitations, 
of which one or more carries its spin and one or more carries its charge. Consequently, the ARPES spectral function does not have a pole 
contribution, but rather consists of a multi-particle continuum. As it has been detailed in the method section of this article, 
the exploration of the shape of energy distribution curves (EDC) and
momentum distribution curves (MDC) plots shows clearly the 
enlargement of the lifetime, which can be originated by a spin-charge separation. This approximative evaluation is just 
proportional to various interaction strengths present in the system. \\ \\
{\bf Supplementary Note 3: Definition and validity of some quantities used in the theoretical analysis}\\ 

On the one hand, the distributions $2t\,\eta_{\rm c} (\Lambda)$ and $2t\,\eta_{\rm s} (\Lambda)$ appearing in the expressions of
the pseudofermion energy dispersions $\varepsilon_{\rm c} (q)$ and $\varepsilon_{\rm s} (q')$
given in Eq. (8) are the unique solutions of the coupled integral equations provided in the supplementary Equations (\ref{ceta})
and (\ref{Jeta}) where $u=U/4t>0$.

On the other hand, the $q$ and $q'$ dependence of the dispersions $\varepsilon_{\rm c} (q)$ and $ \varepsilon_{\rm s} (q')$
occurs through that of the momentum rapidity function $k = k (q)$ for $q \in [-\pi,\pi]$ and
spin rapidity function $\Lambda = \Lambda (q')$ for $q' \in [-k_{\rm F},k_{\rm F}]$, respectively.
Those are defined in terms of their inverse functions $q = q (k)$ for $k \in [-\pi,\pi]$ and
$q' = q' (\Lambda)$ for $\Lambda \in [-\infty,\infty]$ given in the supplementary Equations (\ref{eq-cont1}) and
(\ref{eq-cont}), respectively. The distributions $2\pi\rho (k)$ and $2\pi\sigma (\Lambda)$ in these equations are the
unique solutions of the coupled integral equations provided in the supplementary Equations (\ref{rho}) and (\ref{sigma}).

Within the $\beta =c,s$ pseudofermion representation, the conformal-field theory dressed-charge matrix 
and the corresponding matrix $(Z^{-1})^T $ 
can be expressed in terms of pseudofermion phase shifts in units of $2\pi$, as given in the supplementary
Equation (\ref{ZZPS}). For the zero-spin density case considered in the article, such matrices can
be written as in the supplementary Equation (\ref{ZZIT}).

In the zero spin-density case, the spin $SU(2)$ symmetry implies that the parameter $\xi_{\rm s}$ appearing 
in supplementary Equation (\ref{ZZIT}) is $u$ independent and given by $\xi_{\rm s} = \sqrt{2}$. The parameter $\xi_{\rm c}$ in 
that equation in turn reads $\xi_{\rm c} = f (\sin Q/u)$ where the function
$f (r)$ is the unique solution of the integral equation given in the supplementary Equation (\ref{xic}).
In it, $Q$ is the parameter defined in the supplementary Equation (\ref{eq-cont1}) and the
function $D (r)$ is provided in the supplementary Equation (\ref{Ar}) where $\Gamma (x)$ is the usual $\Gamma$ function. 

The rapidity phase shifts $2\pi\bar{\Phi }_{\rm c,c} (r,r')$ and $2\pi\bar{\Phi }_{\rm c,s} (r,r')$ 
in the expressions of the $c$ pseudofermion phase shifts $2\pi\Phi_{\rm c,c}\left(q,q'\right) = 2\pi\bar{\Phi }_{\rm c,c} \left(\sin k (q)/u,\sin k (q')/u\right)$
and $2\pi\Phi_{\rm c,s}\left(q,q'\right) = 2\pi\bar{\Phi }_{\rm c,s} \left(\sin k (q)/u,\Lambda (q')/u\right)$
considered in the Methods section are the unique solutions of the integral equations given in the
supplementary Equations (\ref{Phiccm0}) and (\ref{Phicsm0}) where $D (r)$
and $D_0 (r)$ are defined in supplementary Equations (\ref{Ar}) and (\ref{Br}), respectively.

The related ${\rm s}$ pseudofermion phase shifts $2\pi\Phi_{\rm s,c}(\iota k_{\rm F},q)$
and $2\pi\Phi_{\rm s,s}(\iota k_{\rm F},q')$ are provided in the supplementary Equations (\ref{PhiscUinfqFu})
and (\ref{PhissUinfqFu}), respectively. (The corresponding more general ${\rm s}$ pseudofermion phase shifts $2\pi\Phi_{\rm s,s}(q',q)$ and 
$2\pi\Phi_{\rm s,c}(q',q)$ are uniquely defined by the solution of integral equations that are not given here.)

Concerning the validity of the spectral functions expressions in Eqs. (9) and (18), when for an electron removal 
spectral function $\gamma$ branch line there is for 
$(\omega_{\gamma} (k)-\omega)<0$ (ii) no spectral weight and (ii) a very small amount of weight, they are (i) exact and
(ii) a very good approximation for an energy window corresponding to a small energy deviation
$(\omega_{\gamma} (k)-\omega)>0$ from the high-energy branch-line spectrum $\omega_{\gamma} (k)$.
In the present case, the electron removal spectral function expressions are
exact for the ${\rm s}$ branch line and a very good approximation for the $k$ ranges of the ${\rm c}$ and ${\rm c'}$ branch lines for which 
the corresponding power-law exponents are negative, respectively.\\ \\

{\bf Supplementary Note 4: The electron finite-range interactions versus onsite interactions only}\\ 

Can the approach used in the paper to account for the effects of the electron finite-range interactions 
be used to extract information beyond that given by the conventional 1D Hubbard model about the physics of quasi-1D metals?
For each finite effective $U/t$ and electronic density $n<1$, the value of the ${\rm c'}$ branch-line exponent
${\tilde{\zeta}}_{\rm c'} (k)$ in Eq. (19) decreases upon increasing $\alpha$ 
within the range $\alpha \in [\alpha_0,1/8]$ and increases upon further increasing 
it within the interval $\alpha \in ]1/8,\alpha_{\rm max}]$. Consistently, there is a $\alpha$ value 
belonging to the interval $\alpha \in ]1/8,\alpha_{\rm max}]$ for which the ${\rm c'}$ branch line momentum width 
where the exponent ${\tilde{\zeta}}_{\rm c'} (k)$ is negative exact equals that obtained for
the conventional 1D Hubbard model at $\alpha = \alpha_0$. Within the renormalized pseudofermion
dynamical theory introduced
in the paper, this may be behind the agreement reported in
Refs. [\cite{Carmelo-06,Carmelo-08}] of the spectral features predicted by that model pseudofermion
dynamical theory with those observed in 
the quasi-1D metal tetrathiafulvalene tetracyanoquinodimethane (TTF-TCNQ) corresponding to the the 
TCNQ molecular chains\cite{Sing-03,Claessen-02}. A RPDT description would imply
a $K_{\rm c}<1/2$ TLL charge parameter for such chains. By using the conventional 1D Hubbard model, 
the studies of Refs. [\cite{Carmelo-06,Carmelo-08}] rather assumed that $K_{\rm c}>1/2$ 
for the TCNQ molecular chains.

For the 1D Hubbard model with electron finite-range interactions corresponding to $\alpha = 0.78$,
the theoretical ${\rm c}$ branch line spectrum $\omega_{\rm c} (k) = \varepsilon_{\rm c} (\vert k\vert + k_{\rm F})$, 
${\rm c'}$ branch line spectrum $\omega_{\rm c'} (k) = \varepsilon_{\rm c} (\vert k\vert - k_{\rm F})$, 
and ${\rm s}$ branch line spectrum $\omega_{\rm s} (k) = \varepsilon_{\rm s} (k)$ are plotted in Fig. 5 (d) 
for $U/t=0.8$, $t=0.58$\,eV and $n=2/3$. The corresponding branch-line spectra of the 
conventional 1D Hubbard model are plotted for these values in Supplementary Figure 5. 
The ${\rm c}$, ${\rm c'}$, and ${\rm s}$ branch lines appear in the latter figure as full and dashed lines for 
$k$ ranges for which the corresponding exponents are negative and positive, respectively. 

The difference between Fig. 5 (d) and Supplementary Figure 5 refers mainly to the momentum range for which the ${\rm c'}$ branch line 
exponent is negative, which is inexistent for the 1D Hubbard model with finite-range
interactions corresponding to $\alpha =0.78$. An additional difference
is that for the latter model the ${\rm s}$ branch line exponent becomes positive
for $k$ values near the $\pm k_{\rm F}$ Fermi points.\\ \\
{\bf Supplementary Equations}

\begin{equation}
2t\,\eta_{\rm c} (k)  = 2t\sin k + \frac{\cos k}{\pi\,u} \int_{-\infty}^{\infty}d\Lambda\,{2t\,\eta_{\rm s} (\Lambda)\over 1 +  \left({\sin k - \Lambda\over u}\right)^2} \, .
\hspace{5.6cm}{\rm Supplementary}\hspace{0.2cm}{\rm Equation}
\label{ceta}
\end{equation}
\begin{equation}
2t\,\eta_{\rm s} (\Lambda) = {1\over\pi\,u}\int_{-Q}^{Q}dk\,{2t\,\eta_{\rm c} (k)\over 1 +  \left({\Lambda-\sin k\over u}\right)^2} 
- \frac{1}{2\pi\,u} \int_{-\infty}^{\infty}d\Lambda^{\prime}\,{2t\,\eta_{\rm s} (\Lambda^{\prime})\over 1 +  \left({\Lambda -
\Lambda^{\prime}\over 2u})\right)^2} \, . \hspace{2.7cm}{\rm Supplementary}\hspace{0.2cm}{\rm Equation}
\label{Jeta}
\end{equation}
\begin{eqnarray}
q (k) & = & k + \frac{1}{\pi} \int_{-\infty}^{\infty}d\Lambda\,2\pi\sigma (\Lambda)\, \arctan \left({\sin k -
\Lambda\over u}\right)\hspace{0.35cm}{\rm for}\hspace{0.25cm}k \in [-\pi,\pi] \, ,
\nonumber \\
& & q (\pm Q) = \pm 2k_{\rm F} \, ; \hspace{1.00cm} \pm Q = k (\pm 2k_{\rm F}) \, ,
\nonumber \\
& & q (\pm\pi) = \pm\pi \, ; \hspace{1.00cm} \pm\pi = k (\pm\pi) \, . \hspace{6.2cm}{\rm Supplementary}\hspace{0.2cm}{\rm Equation}
\label{eq-cont1}
\end{eqnarray}
\begin{eqnarray}
q' (\Lambda) & = & {1\over\pi}\int_{-Q}^{Q}dk\,2\pi\rho (k)\, \arctan \left({\Lambda-\sin k\over u}\right) 
\nonumber \\
& - & \frac{1}{\pi} \int_{-\infty}^{\infty}d\Lambda^{\prime}\,2\pi\sigma (\Lambda^{\prime})\, \arctan \left({\Lambda -
\Lambda^{\prime}\over 2u}\right)\hspace{0.35cm}{\rm for}\hspace{0.25cm}\Lambda \in [-\infty,\infty] \, ,
\nonumber \\
& & q' (\pm\infty) = \pm k_{\rm F} \, , \hspace{1.00cm} \pm\infty = \Lambda (\pm k_{\rm F}) \, .
\hspace{5.3cm}{\rm Supplementary}\hspace{0.2cm}{\rm Equation}
\label{eq-cont}
\end{eqnarray}
\begin{equation}
2\pi\rho (k) = 1 + \frac{\cos k}{\pi\,u} \int_{-\infty}^{\infty}d\Lambda\,{2\pi\sigma (\Lambda)\over 1 +  \left({\sin k - \Lambda\over u}\right)^2} \, .
\hspace{6.7cm}{\rm Supplementary}\hspace{0.2cm}{\rm Equation}
\label{rho}
\end{equation}
\begin{equation}
2\pi\sigma (\Lambda) = {1\over\pi\,u}\int_{-Q}^{Q}dk\,{2\pi\rho (k)\over 1 +  \left({\Lambda-\sin k\over u}\right) ^2} 
- \frac{1}{2\pi\,u} \int_{-\infty}^{\infty}d\Lambda^{\prime}\,{2\pi\sigma (\Lambda^{\prime})\over 1 +  \left({\Lambda -
\Lambda^{\prime}\over 2u})\right)^2} \, . \hspace{2.8cm}{\rm Supplementary}\hspace{0.2cm}{\rm Equation}
\label{sigma}
\end{equation}
\begin{eqnarray}
Z & = & \left[\begin{array}{cc}
1 & 0 \\
0 & 1
\end{array}\right] +
\sum_{\iota=\pm 1}\left[\begin{array}{cc}
\Phi_{\rm c,c}(\iota 2k_{\rm F},2k_{\rm F}) & \Phi_{\rm c,s}(\iota 2k_{\rm F},k_{\rm F}) \\
\Phi_{s,c}(\iota k_{\rm F},2k_{\rm F}) & \Phi_{s,s}(\iota k_{\rm F},k_{\rm F})  
\end{array}\right] \, ,
\nonumber \\
(Z^{-1})^T & = & \left[\begin{array}{cc}
1 & 0 \\
0 & 1
\end{array}\right] +
\sum_{\iota=\pm 1}(\iota)\left[\begin{array}{cc}
\Phi_{\rm c,c}(\iota 2k_{\rm F},2k_{\rm F}) & \Phi_{\rm c,s}(\iota 2k_{\rm F},k_{\rm F}) \\
\Phi_{s,c}(\iota k_{\rm F},2k_{\rm F}) & \Phi_{s,s}(\iota k_{\rm F},k_{\rm F})  
\end{array}\right]  \, . \hspace{2.8cm}{\rm Supplementary}\hspace{0.2cm}{\rm Equation}
\label{ZZPS}
\end{eqnarray}
\begin{equation}
Z = \left[\begin{array}{cc}
\xi_{\rm c} & \xi_{\rm c}/2 \\
0 & 1/\xi_{\rm s} 
\end{array}\right]
\, ; \hspace{0.5cm}
(Z^{-1})^T = \left[\begin{array}{cc}
1/\xi_{\rm c} & 0 \\
-\xi_{\rm s}/2 & \xi_{\rm s}
\end{array}\right] \, ; \hspace{0.5cm} \xi_{\beta} = \sqrt{2K_{\beta}}\hspace{0.35cm}{\rm for}\hspace{0.25cm}\beta = c,s \, .
\hspace{1.4cm}{\rm Supplementary}\hspace{0.2cm}{\rm Equation}
\label{ZZIT}
\end{equation}
\begin{equation}
f (r) = 1 + \int_{-sin Q\over u}^{sin Q\over u} d r' D (r-r')\,f (r') \, .
\hspace{7.5cm}{\rm Supplementary}\hspace{0.2cm}{\rm Equation}
\label{xic}
\end{equation}
\begin{equation}
D (r) = {1\over\pi}\int_{0}^{\infty} d\omega{\cos (\omega\,r)\over 1+e^{2\omega}} =
{i\over 2\pi} {d\over dr}\ln {\Gamma \Bigl({1\over 2}+i{r\over 4}\Bigr)\,\Gamma
\Bigl(1-i{r\over 4}\Bigr)\over \Gamma \Bigl({1\over 2}-i{r\over 4}\Bigr)\,\Gamma
\Bigl(1+i{r\over 4}\Bigr)} \, . \hspace{3.7cm}{\rm Supplementary}\hspace{0.2cm}{\rm Equation}
\label{Ar}
\end{equation}
\begin{equation}
2\pi{\bar{\Phi }}_{\rm c,c}(r,r') = - D_0 (r-r') 
+ \int_{-{\sin Q\over u}}^{{\sin Q\over u}}dr''\,D (r-r'')\,2\pi{\bar{\Phi }}_{\rm c,c}(r'',r') \, .
\hspace{3.2cm}{\rm Supplementary}\hspace{0.2cm}{\rm Equation}
\label{Phiccm0}
\end{equation}
\begin{equation}
2\pi{\bar{\Phi }}_{\rm c,s}(r,r') = - \arctan\Bigl(\sinh\Bigl({\pi\over 2}(r-r')\Bigr)\Bigr) 
+ \int_{-{\sin Q\over u}}^{{\sin Q\over u}}dr''\,D (r-r'')\,2\pi{\bar{\Phi}}_{\rm c,s}(r'',r') \, .
\hspace{1cm}{\rm Supplementary}\hspace{0.2cm}{\rm Equation}
\label{Phicsm0}
\end{equation}
\begin{equation}
D_0 (r) = 2\int_{0}^{\infty} d\omega{\sin (\omega\,r)\over \omega (1+e^{2\omega})} =
i\,\ln {\Gamma \Bigl({1\over 2}+i{r\over 4}\Bigr)\,\Gamma
\Bigl(1-i{r\over 4}\Bigr)\over \Gamma \Bigl({1\over 2}-i{r\over 4}\Bigr)\,\Gamma
\Bigl(1+i{r\over 4}\Bigr)} \, . \hspace{4cm}{\rm Supplementary}\hspace{0.2cm}{\rm Equation}
\label{Br}
\end{equation}
\begin{equation}
2\pi\Phi_{\rm s,c}(\iota k_{\rm F},q) = -{\iota\,\pi \over \sqrt{2}}\hspace{0.15cm}{\rm for}\hspace{0.15cm}\iota = \pm 1 \, .
\hspace{7.7cm}{\rm Supplementary}\hspace{0.2cm}{\rm Equation}
\label{PhiscUinfqFu}
\end{equation}
\begin{eqnarray}
& & 2\pi\Phi_{\rm s,s}(\iota k_{\rm F},q') = \iota\,\pi {(\xi_{\rm s} - 1)(\xi_{\rm s} + (-1)^{\delta_{q',\iota k_{\rm F} }})\over \xi_{\rm s}}
\nonumber \\
& & = {\iota\,\pi\over \sqrt{2}} (\sqrt{2} - 1)(\sqrt{2} + (-1)^{\delta_{q',\iota k_{\rm F} }})\hspace{0.15cm}{\rm for}\hspace{0.15cm}\iota = \pm 1 \, . 
\hspace{5cm}{\rm Supplementary}\hspace{0.2cm}{\rm Equation}
\label{PhissUinfqFu}
\end{eqnarray}

%%%%%%%%%%%%%%%%%%%%%%%%%%%%%%%%%%%%%%%%%%%%%%%%%%%%%%%%%%%%%%%%%%%%%%%%%%

%%%%%%%%%%%%%%%%%%%%%%%%%%%%%%%%%%%%%%%%%%%%%%%%%%%%%%%%%%%%%%%%%%%%%%%%%%

\clearpage

\begin{figS}
%\begin{center}
\centerline{\includegraphics[width=10cm]{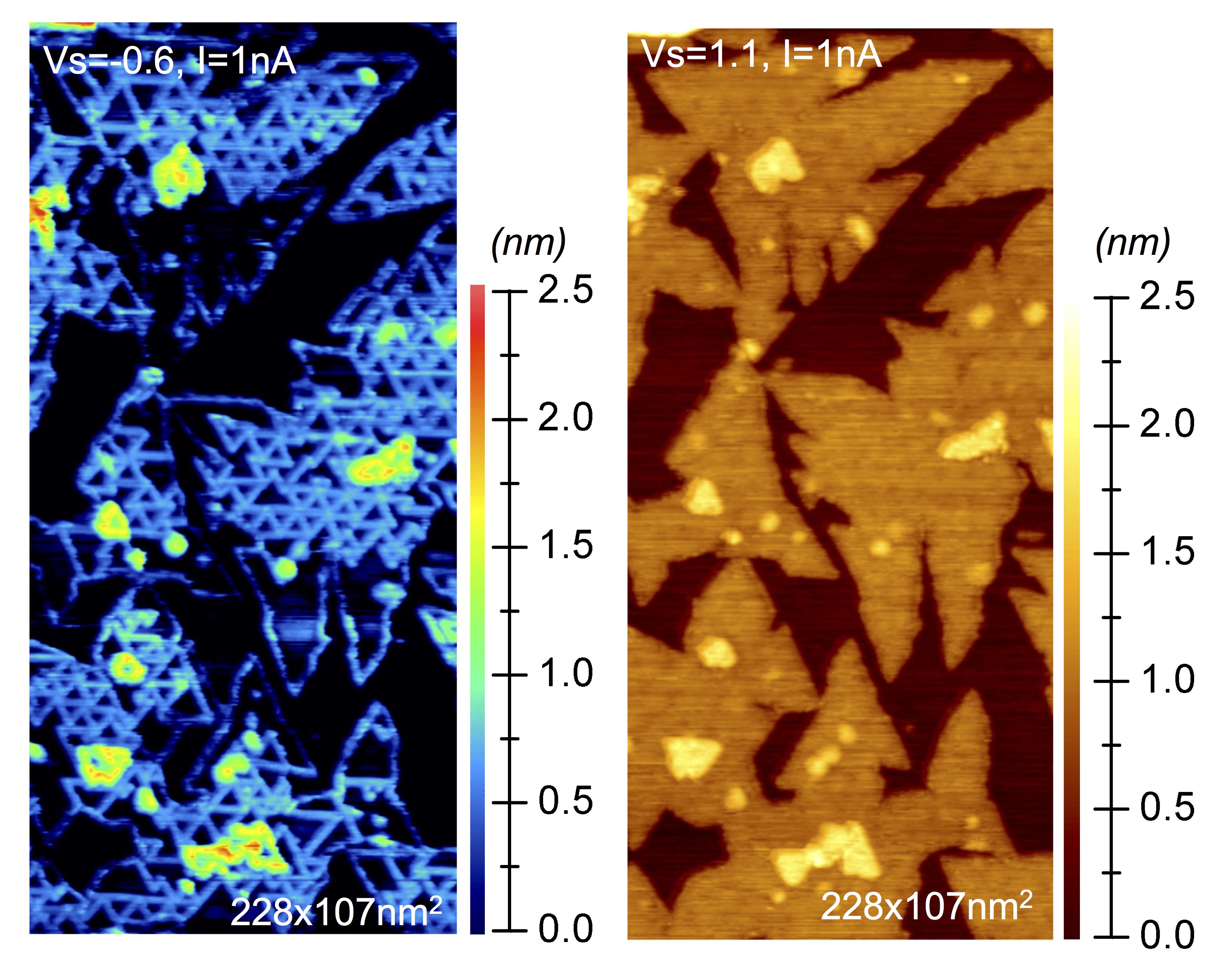}}
\caption{Supplementary Fig. 1.  {\bf Scanning tunnelling micrographs of submonolayer MoSe$_2$.}  The imaged islands are
predominantly monolayer thick MoSe$_2$ flakes, which have been used in ARPES measurements. The STM data are recorded of the same surface area with different tunnelling conditions are shown. One image emphasizes the metallic MTBS and island edges, the other tunnelling conditions emphasizes the island shape and topography. tunnelling conditions are indicated in the figures.}
\label{figureS1}
%\end{center}
\end{figure}

\begin{figS}
%\begin{center}
\centerline{\includegraphics[width=12.00cm]{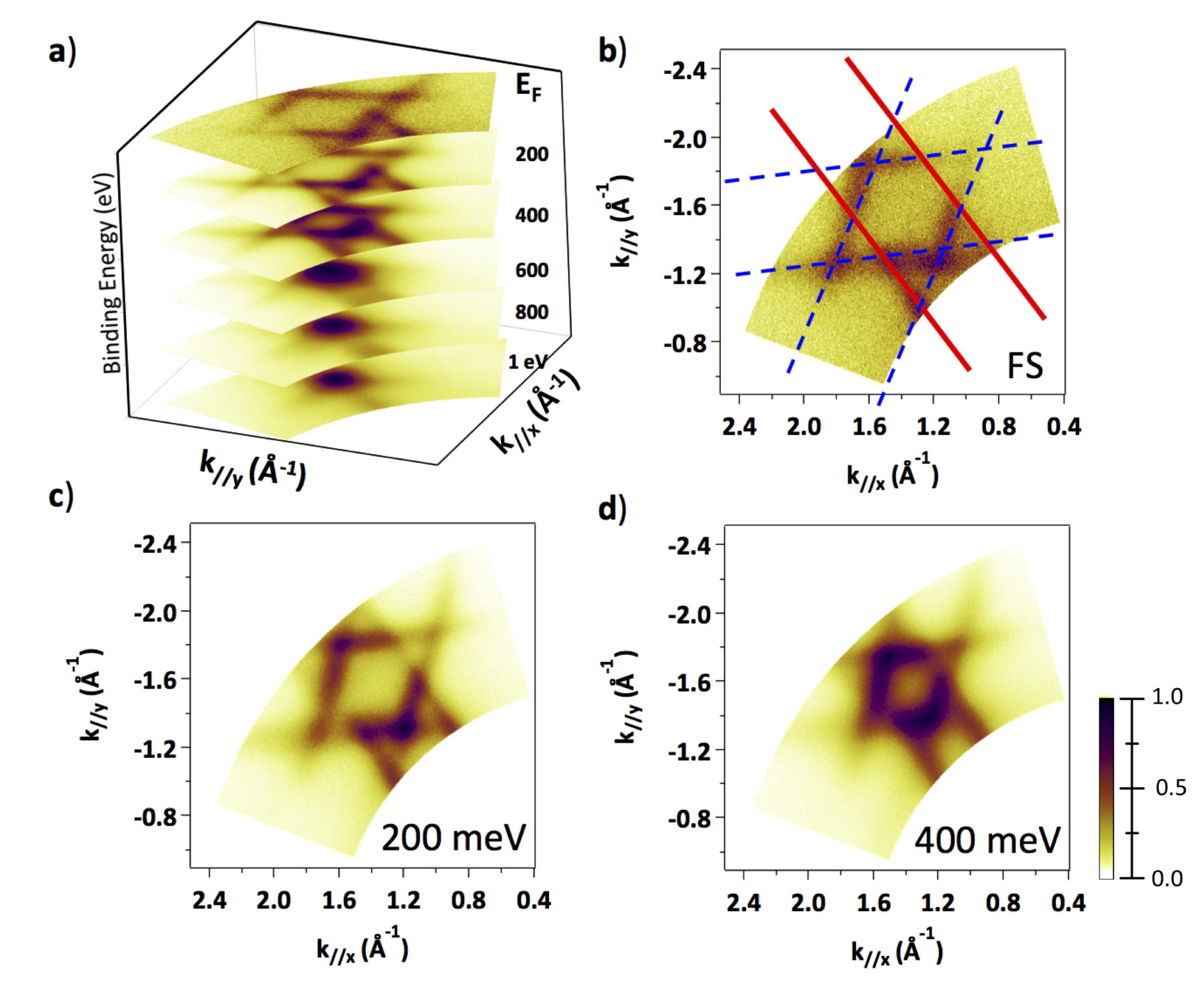}}
\caption{Supplementary Fig. 2. {\bf Fermi-surface and other constant energy surfaces of MTBs.}  Experimental measurements of a constant energy surfaces at different binding energies are represented in a 3D plot in (a).  Spectral function around the $\Gamma$-point at the Fermi level and binding energies of 200 meV and 400 meV are plotted in (b) (c) and (d), respectively. The data shown were obtained with circular polarized light and the intensity for left and right polarized light are added together. The measurements are close to the 4-point of the 2nd BZ of the MoSe$_2$ unit cell.}
\label{figureS2}
%\end{center}
\end{figure}
		
\begin{figS}
%\begin{center}
\centerline{\includegraphics[width=10.00cm]{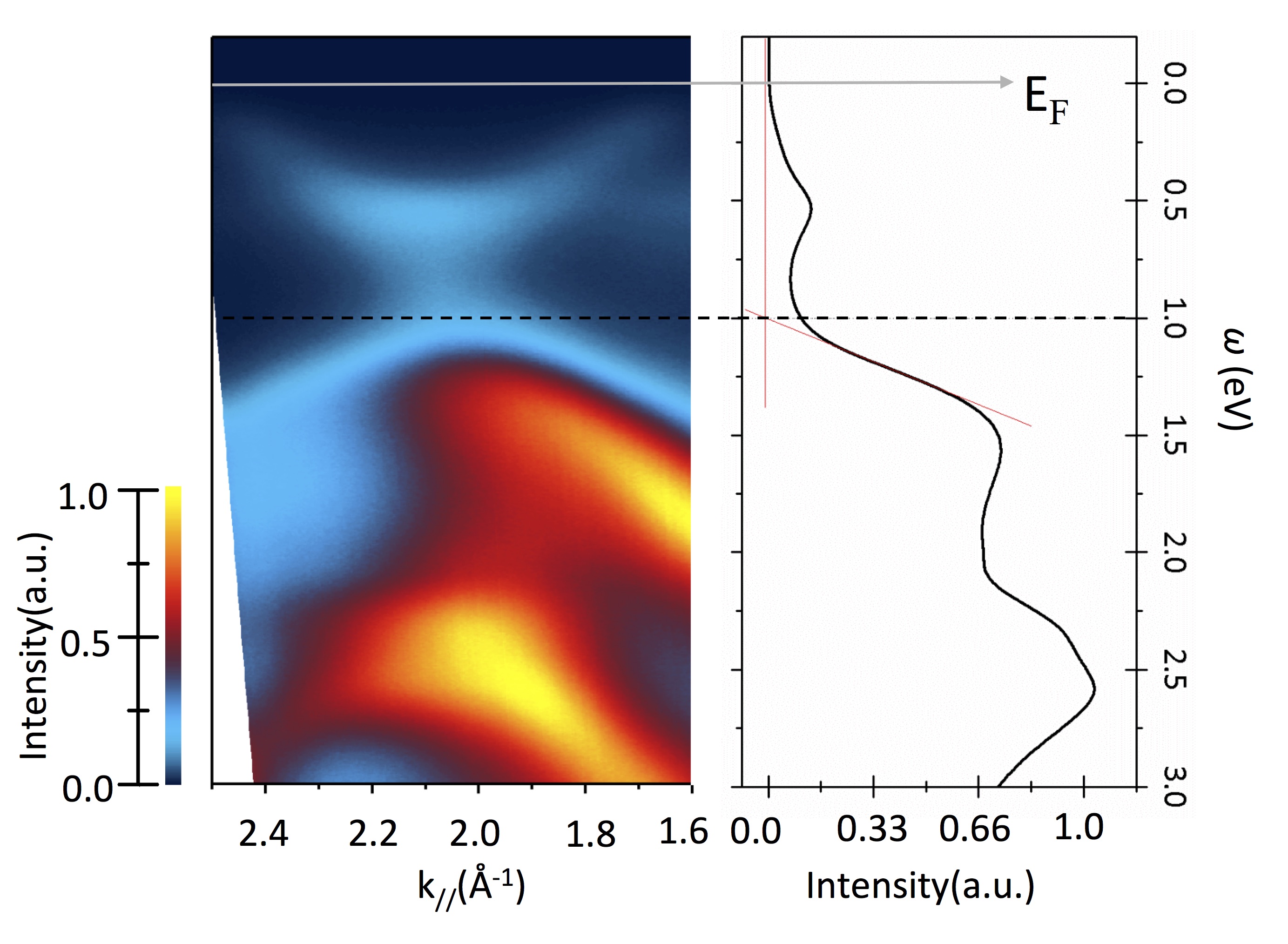}}
\caption{Supplementary Fig. 3. {\bf Gap and top of the valence band of MoSe$_2$ flakes with line defects} Wide energy range photoemission spectrum around Gamma point showing the 1DES and the valence band of the MoSe$_2$ host material. The 1DES lies entirely within the band gap of the semiconducting MoSe$_2$, thus no overlap of electronic states are observed that could affect the interpretation of the 1DES.}
\label{figureS3}
%\end{center}
\end{figure}	
				
\begin{figS}
%\begin{center}
\centerline{\includegraphics[width=12.00cm]{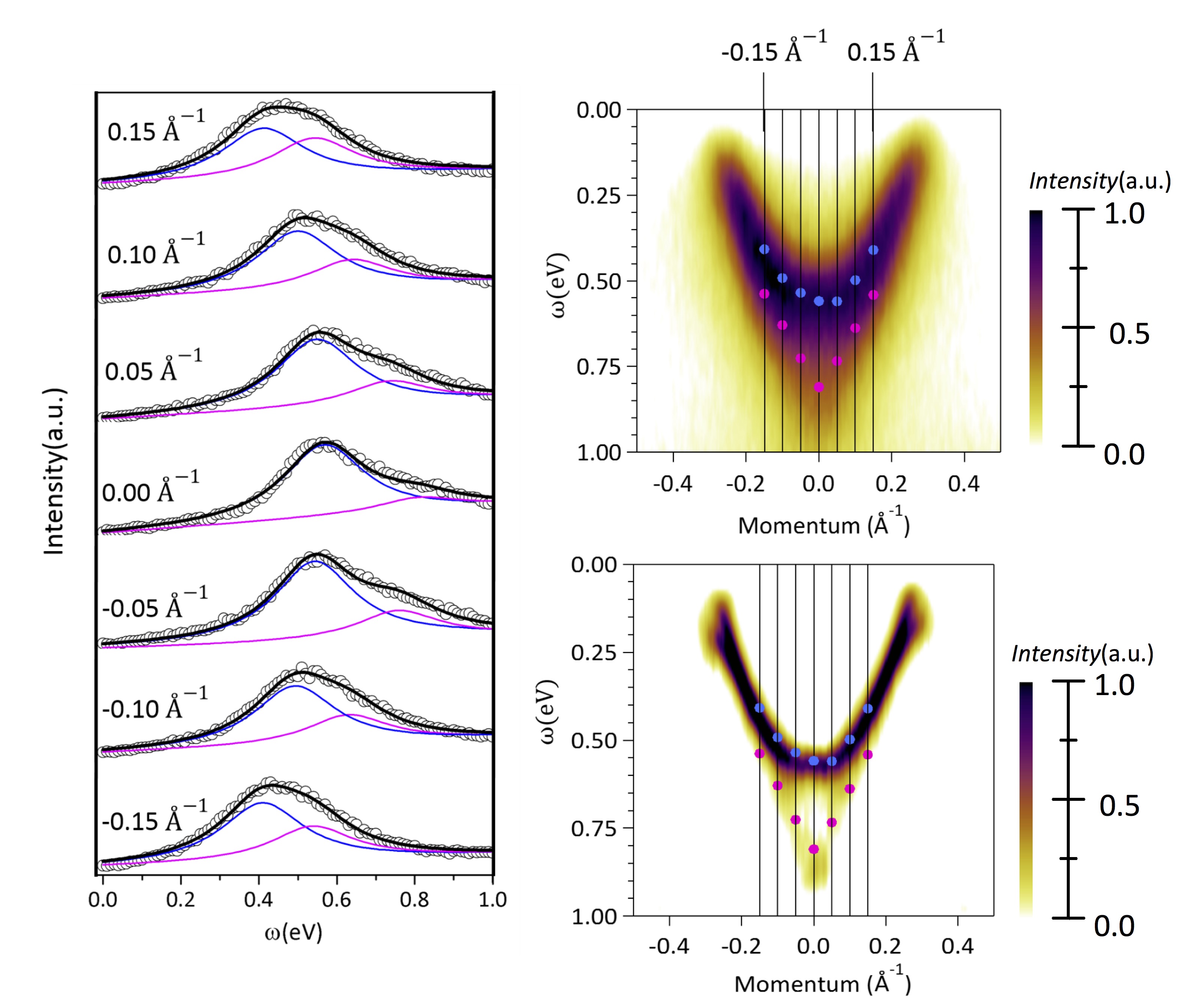}}
\caption{Supplementary Fig. 4. {\bf Selected energy distribution curves (EDCs).} The EDCs contain spectral weight that cannot be fit with a single component. A fit using two components with a Lorentzian line shape and by assuming a linear background is shown on the left. The fitted peak positions are shown on the right superimposed on the spectral intensity map as well as on the processed data that shows the cusps in the second derivative of the data. The peak positions determined from fitting with two peaks agree well with the band dispersion obtained by calculating the second derivatives of the data.}
\label{figureS4}
%\end{center}
\end{figure}

\begin{figS}
\centerline{\includegraphics[width=9.00cm]{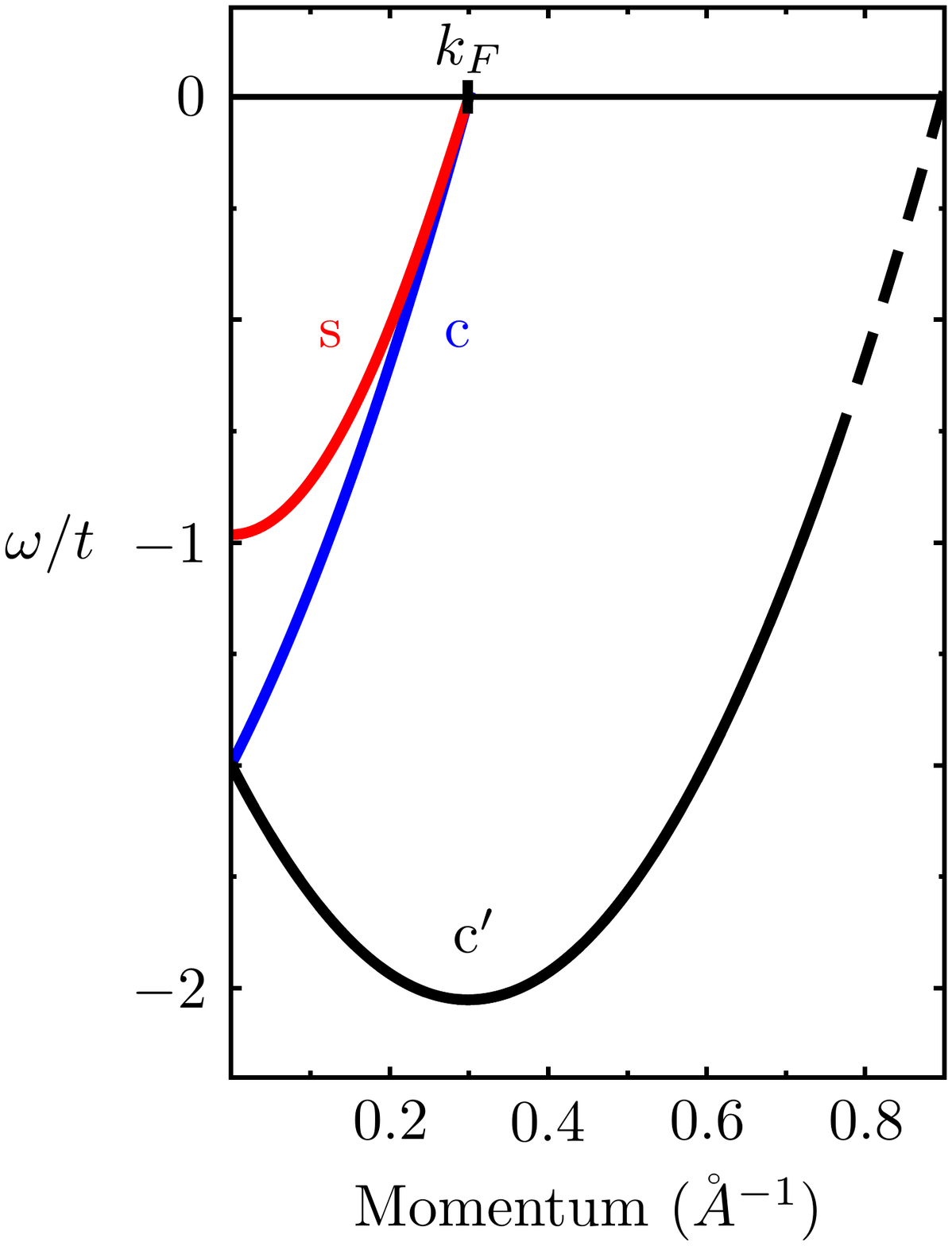}}
%\centerline{\includegraphics[scale=0.5]{FigS5}}
\caption{Supplementary Fig. 5. {\bf Conventional Hubbard model branch-lines spectra.}
The same ${\rm c}$, ${\rm c'}$ and ${\rm s}$ branch line spectra as in Fig. 5 (d) now for the conventional 
1D Hubbard model with the same $U/t=0.8$, $t=0.58$\,eV and $n=2/3$ values.
The full and dashed lines refer to momentum ranges for which the corresponding exponents are negative 
and positive, respectively. The differences relative to Fig. 5 (d) lay in the momentum range 
for which the ${\rm c'}$ branch line exponent is negative, which is absent in the case of
the model renormalised by finite-range interactions corresponding to $\alpha = 0.78$, and the ${\rm s}$ branch line exponent 
of the conventional 1D Hubbard model being negative for its whole momentum range.}
\label{figureS5}
\end{figure}